\documentclass[twocolumn,showpacs,preprintnumbers,amsmath,amssymb,prl,superscriptaddress]{revtex4-1}
\usepackage{bbm}
\usepackage{mathrsfs}
\usepackage{graphicx}
\usepackage{dcolumn}
\usepackage{bm}
\usepackage{amsmath}
\usepackage{amsfonts}
\usepackage{color}
\usepackage[colorlinks=true,citecolor=blue,anchorcolor=blue]{hyperref}
\usepackage{floatrow}

\begin{document}

\title{Distribution of Conductances in Chiral Topological Superconductor Junctions}
\author{Biao Lian}
\thanks{biao@princeton.edu}
\affiliation{Princeton Center for Theoretical Science, Princeton University, Princeton, New Jersey 08544, USA}
\author{Jing Wang}
\thanks{wjingphys@fudan.edu.cn}
\affiliation{State Key Laboratory of Surface Physics, Department of Physics, Fudan University, Shanghai 200433, China}
\affiliation{Collaborative Innovation Center of Advanced Microstructures, Nanjing 210093, China}

\begin{abstract}
We study the electronic transport of heterojunctions made of chiral topological superconductors (TSCs) with $N$ chiral Majorana fermion modes and quantum Hall insulators (QHIs) with integer Chern number $C$. In the weak disorder regime, we show the two-terminal conductance $\sigma_{12}$ of a QHI-TSC-QHI junction is generically non-quantized, but obeys a certain distribution determined by $C$ and $N$, which is induced by random SO($N$) rotations of chiral Majorana fermion mode basis on the TSC edges. Oppositely, in the strong disorder regime, $\sigma_{12}$ tends to be a quantized value. We conclude with a brief discussion on the fractionally quantized thermal conductances of the junction.
\end{abstract}

\date{\today}

%\pacs{
%        73.22.-f  % Electronic structure of nanoscale materials and related systems
%        02.20.-a  % Group theory
%        73.43.-f  % Quantum Hall effects
%      }

\maketitle

%\emph{Introduction.}
As one of the simplest topological states of matter, the two-dimensional (2D) chiral topological superconductor (TSC) has attracted lots of attentions \cite{moore1991,read2000,ivanov2001,mackenzie2003,kitaev2006,fu2008,sau2010,alicea2010,qi2009}. It is characterized by the Bogoliubov-de Gennes (BdG) Chern number~\cite{schnyder2008,kitaev2009}, which take integer values $N\in\mathbb{Z}$ and leads to $N$ chiral Majorana fermion modes (CMFMs) propagating on its edge. The $N=1$ TSC, also known as the $p_x+ip_y$ TSC \cite{read2000}, is extensively studied and has been experimentally realized recently~\cite{he2017}. The understanding of CMFM edge conduction is incomplete due to their charge neutral nature. In analogy of the chiral quantum Hall (QH) edge modes, a fruitful approach is to study scattering between those chiral QH modes by measuring transport through junctions between regions of different carrier density~\cite{williams2007,kim2007,amet2014}. In particular, the single CMFM backscattering in a $N=1$ TSC junction with two Chern number $C=1$ QH insulators (QHIs) is shown to exhibit to a half quantized electrical conductance $e^2/2h$~\cite{qi2010b,chung2011,wang2015c,he2017,lian2017,lian2018}. However, the electronic transport of TSC junctions with generic BdG Chern number $N$ is still unknown.

%In general, the TSC with an even BdG Chern number $N$ can be realized by QHI with Chern number $N/2$ \cite{jiang2018} under $s$-wave superconducting proximity, while the TSC with higher odd $N$ can also be implemented in materials \cite{wang2018}.

In this Letter, we study the electrical conductance of a QHI-TSC-QHI junction with generic QHI Chern number $C$ and TSC BdG Chern number $N$. Different from the $|N|=1$ case, most $|N|>1$ junctions do not exhibit a quantized conductance. Instead, we show the conductance of the junction obeys a certain distribution among all samples and physical conditions, which is a unique function of $(C,N)$ in the weak disorder limit. The conductance distribution results from the random SO($N$) rotations of CMFM basis on the TSC edges due to disorders. While in the strong disorder limit, we find the junction conductance tends to be a quantized value given by $(C,N)$. We conclude with brief discussions on the thermal conductances of the junction contributed by CMFMs and the possible material realization of such junctions.

%Majorana fermions have attracted intense interest in both particle physics and condensed matter physics~\cite{wilczek2009,elliott2015}. The chiral Majorana fermion, a massless fermionic particle being its own antiparticle, could arise as a one-dimensional (1D) quasiparticle edge states of a 2D topological states of quantum matter~\cite{moore1991,read2000,mackenzie2003,kitaev2006,fu2008,sau2010,alicea2010,qi2009}. The propagating chiral Majorana fermions could lead to non-abelian braiding~\cite{lian2017} and may be useful in topological quantum computation~\cite{kitaev2003,nayak2008}. A simple example hosting chiral Majorana fermion is the $p_x+ip_y$ chiral topological superconductor (TSC) with a Bogoliubov-de Gennes (BdG) Chern number $N=1$, which can be realized from a quantum anomalous Hall (QAH) insulator with a Chern number $C=1$ in proximity to an $s$-wave superconductor~\cite{qi2010b,chung2011,wang2015c}. The quantum transport in a QAH-TSC-QAH junction is predicted to exhibit a half quantized conductance plateau induced by a \emph{single} chiral Majorana fermion~\cite{chung2011,wang2015c,lian2016a}, which has been recently observed in the Cr-doped (Bi,Sb)$_2$Te$_3$ thin film QAH system in proximity with Nb superconductor~\cite{he2017}.

%\emph{Model.}

\begin{figure}[b]
\begin{center}
\includegraphics[width=3.4in]{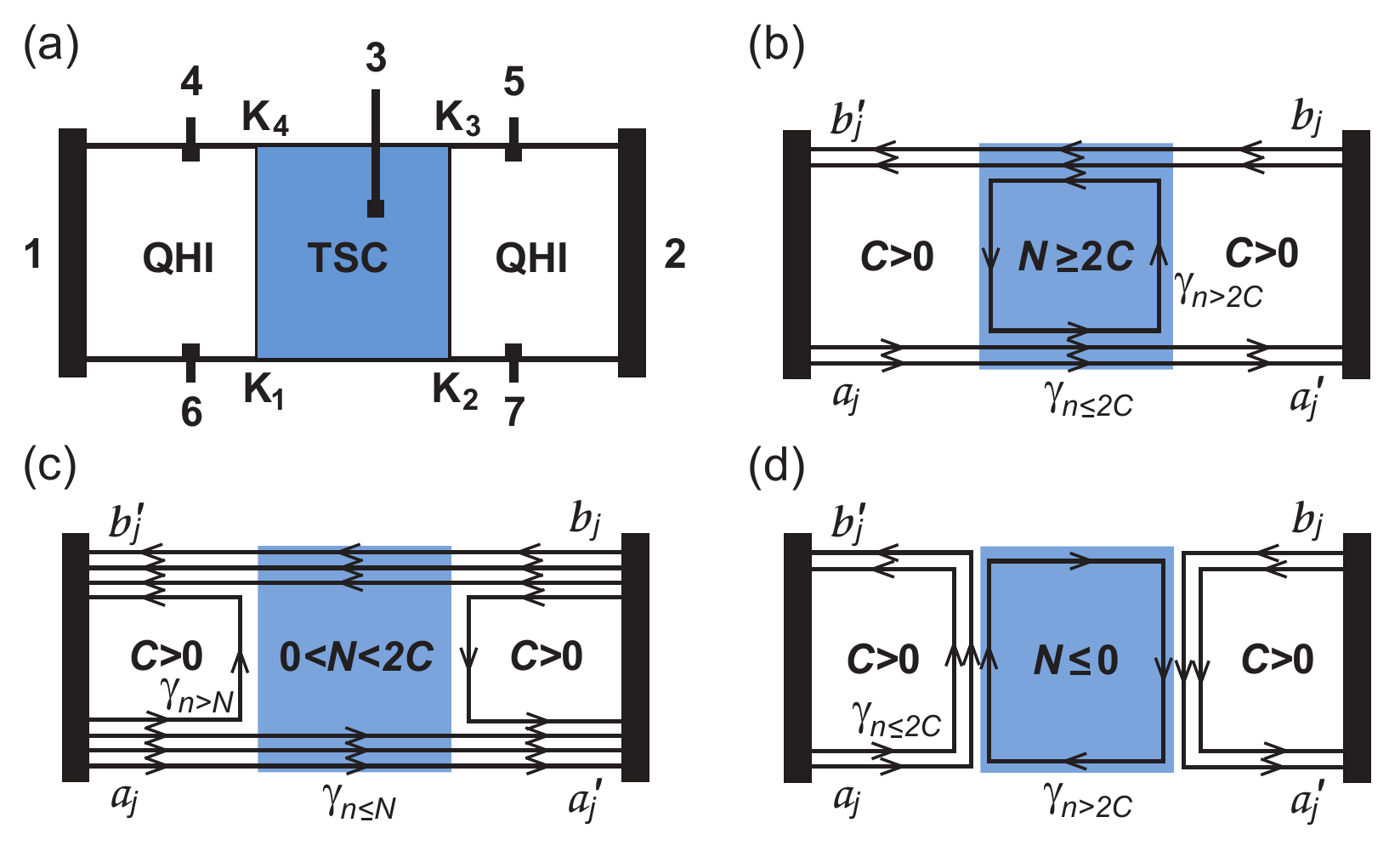}
\end{center}
\caption{(a) The QHI-TSC-QHI junction and the leads for transport measurements. (b-d) Illustration of the CMFMs in junctions with (b) $N\ge2C$, (c) $0<N<2C$, and (d) $N\le0$.}
\label{fig1}
\end{figure}

The geometry of the QHI-TSC-QHI junction we shall study is shown in Fig. \ref{fig1}(a), where we assume the QHIs on the left and right have a Chern number $C>0$, and the chiral TSC in the middle has a BdG Chern number $N$. A number of metallic leads are shown for transport measurements, where lead $3$ connects to the TSC bulk, while all the other leads are on QHI edges. We shall assume the electrical (thermal) currents are only applied on leads $1$, $2$ and $3$, while leads $4$-$7$ are for voltage (temperature) measurements. The electrical conductance of the junction is governed by the generalized Landauer-B\"uttiker formula \cite{anantram1996,entin2008},
\begin{eqnarray}\label{LB}
&&I_1=\frac{e^2}{h}[(C-r+r_A)(V_1-V_{sc})+(t-t_A)(V_{sc}-V_2)],
\nonumber
\\
&&I_2=\frac{e^2}{h}[(C-r+r_A)(V_2-V_{sc})+(t-t_A)(V_{sc}-V_1)],
\nonumber
\\
&&I_3=-I_1-I_2\ ,\quad V_3=V_{sc}\ ,
\end{eqnarray}
Here $I_i$ and $V_i$ are the inflow current and voltage of lead $i$, $V_{sc}$ is the voltage of the TSC, $e$ is the electron charge, $h$ is the Planck constant, and we have further assumed the contact resistance between lead $3$ and the TSC bulk is zero, which is appropriate when lead $3$ is a good metal. The coefficients $r,r_A,t$ and $t_A$ are the normal reflection, Andreev reflection, normal transmission and Andreev transmission coefficients between leads $1$ and $2$, respectively, which satisfy $r+r_A+t+t_A=C$. From Eq. (\ref{LB}), only $r-r_A$ and $t-t_A$ are needed, the two-terminal conductance of the junction is derived as $\sigma_{12}=I/(V_1-V_2)=(C-r+r_A+t-t_A)/2$ in units of $e^2/h$, where the current $I$ is applied between leads $1$ and $2$, i.e., $I=I_1=-I_2$. Besides, one can derive the conductance between leads $1$ and $3$ as $\sigma_{13}=I/(V_1-V_3)=C-r+r_A-\frac{(t-t_A)^2}{C-r+r_A}$, for which the current $I=I_1=-I_3$. The generic Landauer-B\"uttiker formula for all the leads can be found in the Supplementary Material \cite{supple}, from which one can calculate the other resistances $R_{ij,kl}=(V_k-V_l)/I$ with current $I=I_i=-I_j$. In particular, the Hall resistance of the standard Hall bar is always $R_{12,46}=R_{12,57}=1/C$, while the longitudinal resistance $R_{12,45}=R_{12,67}=1/\sigma_{12}-1/C$ in units of $h/e^2$.

There are three cases to be distinguished: $N\ge2C$, $0<N<2C$ and $N\le0$ cases, of which the edge states are schematically shown in Fig. \ref{fig1}(b)-(d), respectively. %which we will study separately in the below.
In all three cases, there are $C$ charged chiral electron modes (CEMs) on the edges between QHIs and the vacuum. For convenience, we denote the annihilation and creation operators of the $j$-th CEM ($1\le j\le C$) on the four QHI edges next to TSC corners $K_{m}$ ($1\le m\le4$, see Fig. \ref{fig1}(a)) as $a_j,a_j',b_j,b_j'$ and $a_j^\dag,a_j'^\dag,b_j^\dag,b_j'^\dag$, respectively. These $C$ CEMs can then be rewritten as $2C$ CMFMs $\gamma_n$ ($1\le n\le2C$), where we choose the CMFM basis so that they are related to the electron basis via $a_j,a_j',b_j,b_j'=(\gamma_{2j-1}+i\gamma_{2j})/\sqrt{2}$ on the corresponding QHI edges, respectively.

%(\textbf{Is this definition correct? $a_j,b_j$ is right, but for $a_j'$, $b_j'$?})

(a) \emph{$N\ge2C$ case.} There are $N-2C$ CMFMs $\gamma_{n}$ ($2C<n\le N$) circulating on the TSC edges $K_4K_1$ and $K_2K_3$, and $N$ CMFMs on the $K_1K_2$ and $K_3K_4$ TSC edges (see Fig. \ref{fig1}(b)). The effective action of the $N$ CMFMs on edge $K_1K_2$ ($K_3K_4$) takes the form
\begin{equation}
S=\int dtdx\left[\sum_{n=1}^{N} i\gamma_n(\partial_t+v_n\partial_x)\gamma_n+i\gamma^TL(x)\gamma\right]\ ,
\end{equation}
where $\gamma=(\gamma_1,\cdots,\gamma_N)^T$ is the Majorana basis, $v_n>0$ is the Fermi velocity of $\gamma_n$, and the function $L(x)$ is an $N\times N$ real antisymmetric matrix. The $L(x)$ term among the $N$ CMFMs is generically contributed by the chemical potential, hopping and pairing on the edge. In the presence of disorders, one has $L(x)=\overline{L}+\delta L(x)$ varying with respect to $x$, where $\overline{L}$ and $\delta L(x)$ are the mean value and random fluctuations, respectively. Via an SO($N$) transformation $\gamma= Q^T(x)\widetilde{\gamma}$ where $Q(x)=\mathcal{P}\exp\left[\int_{-\infty}^xdx'L(x')/\bar{v}\right]$, one can rewrite the action as \cite{levin2007,lee2007,lian2018a}
\begin{equation}\label{Sv}
S=\int dtdx\left[i\widetilde{\gamma}^T(\partial_t-\bar{v}\partial_x)\widetilde{\gamma}+i\widetilde{\gamma}^TQ^T\delta v\partial_x(Q\widetilde{\gamma})\right],
\end{equation}
where $\mathcal{P}$ stands for path ordering, $\bar{v}=\sum_{n}v_n/N$ is the mean Fermi velocity, and $\delta v=\text{diag}(v_1-\bar{v},\cdots,v_n-\bar{v})$ is Fermi velocity anisotropy. Due to the randomness in $L(x)$, the $\delta v$ term is irrelevant and can be ignored \cite{lian2018a,supple}. Therefore, the $N$ CMFMs on the edge $K_1K_2$ ($K_3K_4$) simply undergo an SO($N$) transformation $Q=\mathcal{P}\exp\left[\int_{K_2}^{K_1}dxL(x)/\bar{v}\right]$. Similarly, the $N-2C$ CMFMs on edge $K_2K_3$ ($K_4K_1$) will undergo an SO($N-2C$) transformation after propagation.

%In the presence of disorders, the $L(x)$ term is generically a random function of $x$, and we will assume it is short range correlated with the mean value $\text{Tr}[\overline{L(x)^TL(x')}]=W\delta(x-x')$. The scaling dimension of $W$ can be easily computed to be $d_W=1>0$, so the $L(x)$ term is relevant. To understand the effect of the relevant $L(x)$ term,

The SO($N$) rotation among the CMFMs on the TSC edges will generate an Andreev probability of electron turning into hole, and thus lead to $t-t_A$ and $r-r_A$ between leads $1$ and $2$. We assume the system size is large enough so that the CMFMs lose coherence during their propagations. When an electron mode $a_i$ is incident from lead $1$ and propagates along edge $K_1K_2$, it may arrive at lead $2$ directly as an electron $a_j'$ or hole $a_j'^\dag$, or it may turn into a CMFM $\gamma_{n>2C}$ on edge $K_2K_3$, circulate around the TSC edge for several laps, and then propagate to either lead $1$ or lead $2$ as an electron or hole.
%as $a_j'$ or $a_j'^\dag$, or go back to lead $1$ as $b_j'$ or $b_j'^\dag$.
To compute $r,r_A,t,t_A$, we need to sum over the contributions $r^{(n)},r_A^{(n)},t^{(n)},t_A^{(n)}$ from all the above paths, where $n\ge1$ denotes the times a circulating path passes edge $K_1K_2$ before reaching lead $1$ (for $r$ and $r_A$) or lead $2$ (for $t$ and $t_A$).

However, the calculation of $t-t_A$ and $r-r_A$ can be simplified by the following fact: once an incident electron turns into a CMFM $\gamma_{n>2C}$ on edge $K_2K_3$, it will have equal probabilities on average to become electron or hole afterwards, since $\gamma_{n>2C}$ are particle-hole symmetric and unrelated to the QHI edge charge basis. Therefore, we obtain $r^{(n)}=r_A^{(n)}$ for all $n\ge1$, and $t^{(n)}=t_A^{(n)}$ for all $n\ge2$. This yields $r-r_A=0$, and $t-t_A=t^{(1)}-t_A^{(1)}$ only contributed by the shortest path from lead $1$ to lead $2$ which passes edge $K_1K_2$ once.

The coefficients $t^{(1)}$ and $t_A^{(1)}$ are generically determined by the SO($N$) rotation matrix $Q$ on edge $K_1K_2$. After propagating along edge $K_1K_2$, the electron annihilation operator $a_i$ has a probability $|u_j^\dag Qu_i|^2$ to become $a_j'$, and a probability $|u_j^T Qu_i|^2$ to become $a_j'^\dag$, where $u_j=(0,\cdots,0,1,i,0,\cdots,0)^T/\sqrt{2}$ is the annihilation operators $a_j$ and $a_j'$ written in the CMFM basis (with the $(2j-1)$-th and $2j$-th components being $1/\sqrt{2}$ and $i/\sqrt{2}$, respectively). We thus have
\begin{equation}\label{tta}
t-t_A=t^{(1)}-t_A^{(1)}=\sum_{i,j=1}^{C}\left(|u_j^\dag Qu_i|^2-|u_j^T Qu_i|^2\right)\ .
\end{equation}

%the edge disorders are generically weak compared to the background pairing, hopping and chemical potential, and we have

We first examine the weak disorder limit $|\delta L(x)|\ll |\overline{L}|$. The antisymmetric $L(x)$ can be block diagonalized into $L(x)=iG(x)\Omega(x)G^T(x)$, where $G(x)$ is an SO($N$) matrix, while $\Omega(x)=\text{diag}(\omega_1(x)\sigma_y,\cdots, \omega_{N/2}(x)\sigma_y)$ for even $N$, and $\Omega(x)=\text{diag}(\omega_1(x)\sigma_y,\cdots, \omega_{(N-1)/2}(x)\sigma_y,0)$ for odd $N$, with $\sigma_{x,y,z}$ being the $2\times 2$ Pauli matrices.
Weak disorder implies one can choose $G(x)=\overline{G}+\delta G(x)$ so that $|\delta G(x)|\ll \overline{G}$, and $\omega_l(x)=\overline{\omega_l}+\delta\omega_l(x)$ where $|\delta\omega_l(x)|\ll|\overline{\omega_l}|$ ($1\le\ell\le N/2$). One can then approximate $Q\approx\overline{G} \Lambda\overline{G}^T$, where $\Lambda=\text{diag}(e^{i\phi_1\sigma_y},\cdots,e^{i\phi_{N/2}\sigma_y})$ for even $N$, or $\Lambda=\text{diag}(e^{i\phi_1\sigma_y},\cdots,e^{i\phi_{(N-1)/2}\sigma_y},1)$ for odd $N$, %$\lambda_{2l-1}=e^{i\phi_l}$ and $\lambda_{2l}=e^{-i\phi_l}$ ($l\le N/2$) are conjugate pairs, and $\gamma_N=1$ if $N$ is odd.
and the angles $\phi_l=\int_A^B \omega_{l}(x)dx/\bar{v}$. For convenience, we define $\overline{G}\equiv(\bm{g}_1,\cdots,\bm{g}_N)$, then the vectors $\bm{g}_l=(g^1_l,\cdots,g^N_l)^T$ $(1\le l\le N)$ form an orthonormal frame. Noting that $\overline{G}$ is not unique, but is determined only up to $\lfloor N/2\rfloor$ independent SO(2) rotations in the $(\bm{g}_{2l-1},\bm{g}_{2l})$ planes ($1\le l\le \lfloor N/2\rfloor$), where $\lfloor y\rfloor$ is the floor function of number $y$.
%Physically, $\bm{g}_{2l-1}\pm i\bm{g}_{2l}$ are simply the BdG eigenvectors of the TSC edge.
Therefore, we should identify all equivalent $\overline{G}$, and $\overline{G}$ lives in the coset space SO$(N)/$SO$(2)^{\lfloor N/2\rfloor}$. When the edge length $d\gg2\pi\bar{v}/(\overline{\delta\omega_l^2})^{1/2}$ (which is nothing but the decoherence condition we assumed), $\phi_l$ can be regarded as uniformly random in $[0,2\pi)$. Thus,
for a given $\overline{G}$, the coefficient $t(\overline{G})-t_A(\overline{G})$ is the uniform average of Eq. (\ref{tta}) over all $\phi_{l}$, which can be expressed as \cite{supple}
\begin{equation}\label{ttaG}
t(\overline{G})-t_A(\overline{G})=\sum_{l=1}^{\lfloor N/2\rfloor}\left(\sum_{j=1}^C A_{2l-1,2l}^{2j-1,2j}\right)^2\ ,
\end{equation}
where $A_{lk}^{ij}=g_l^ig_k^j-g_k^ig_l^j$ is the projection of unit area in the $(\bm{g}_i,\bm{g}_j)$ plane onto the $(\gamma_l,\gamma_k)$ plane.
%is antisymmetric under $i\leftrightarrow j$ or $l\leftrightarrow k$.
They satisfy the constraints
\begin{eqnarray}
&&\sum_{l<k}(A_{lk}^{ij})^2=1,\ \ (i\neq j),
\nonumber
\\
&&\sum_{i<j}(A_{lk}^{ij})^2=1,\ \ (l\neq k),
\\
&&\sum_{n}(A_{2l-1,2l}^{in}A_{2k-1,2k}^{nj}-A_{2l-1,2l}^{jn}A_{2k-1,2k}^{ni})=0
\nonumber
\end{eqnarray}
Together with $r-r_A=0$, one can then compute $\sigma_{12}$ and $\sigma_{13}$ for $N\ge2C$ as functions of $\overline{G}$, which generically depends sample and physical conditions.

\begin{figure}[tbp]
\begin{center}
\includegraphics[width=3.4in]{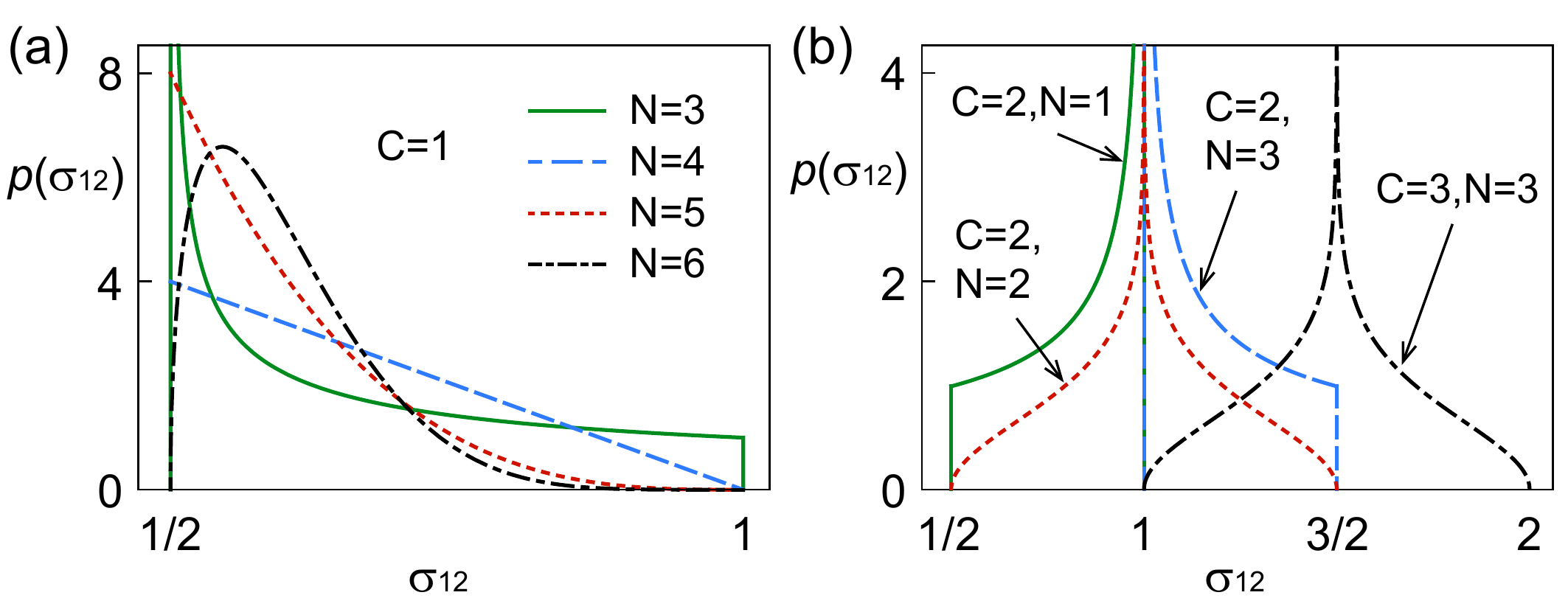}
\end{center}
\caption{The distribution functions $p(\sigma_{12})$ of conductance $\sigma_{12}$ for weakly disordered junctions with (a) $C=1,\ 3\le N\le6$, (b) $C=2,\ 1\le N\le 3$ and $C=3,\ N=3$, respectively.}
\label{fig2}
\end{figure}

%The matrix $\overline{G}$ are generically sample/physical condition dependent, $\sigma_{12}$ and $\sigma_{13}$ as functions of $\overline{G}$ will obey certain distributions.
Next, by assuming that $\overline{G}$ %of various samples and physical conditions are
is uniformly distributed in the coset space SO$(N)/$SO$(2)^{\lfloor N/2\rfloor}$, we can find a specific distribution of $\sigma_{12}$ and $\sigma_{13}$ for various samples/physical conditions. Such an assumption is appropriate given no knowledge about the experimental systems. In particular, for $C=1$ and $N\ge2$, the unnormalized probability distribution of $\sigma_{12}$ is~\cite{supple}
%can be analytically derived to be
\begin{equation}
p(\sigma_{12})=
\begin{cases}
&\delta(\sigma_{12}-1)\ ,\qquad\qquad\qquad (N=2)\\
&\left(\sigma_{12}-\frac{1}{2}\right)^{\alpha_N}\left(1-\sigma_{12}\right)^{\beta_N}\ ,\ (N>2)\\
\end{cases}
\end{equation}
where $\sigma_{12}\in[\frac{1}{2},1]$ in units of $e^2/h$, and the power indices are given by $\alpha_N=\frac{\lfloor N/2\rfloor }{2}-1$ and $\beta_N=\frac{N(N-1)}{4}-\frac{\lfloor N/2\rfloor }{2}-1$. Fig. \ref{fig2}(a) shows the normalized $p(\sigma_{12})$ for $C=1$ and $3\le N\le 6$, and one sees their functional are quite different.

The analytical results of $p(\sigma_{12})$ for generic $C>1$ %(and $N\ge2C$)
are difficult. However, we can still obtain the mean value of $\sigma_{12}$~\cite{supple}. From the geometrical meaning of $A_{kl}^{ij}$, one can show the mean value $\overline{A_{kl}^{ij}A_{k'l'}^{i'j'}}=2\delta_{ii'}\delta_{jj'}\delta_{kk'}\delta_{ll'}/N(N-1)$ for $i<j, k<l, i'<j', k'<l'$. Therefore, the mean value of $\sigma_{12}$ in units of $e^2/h$ is
\begin{equation}\label{conductance1}
\overline{\sigma_{12}}=\frac{C+\overline{(t-t_A)}}{2}=\frac{C}{2}\left[1+\frac{2\lfloor N/2\rfloor}{N(N-1)}\right]\ .
\end{equation}

(b) \emph{$0\le N<2C$ case.} There are $N$ CMFMs $\gamma_{n\le N}$ on edges $K_1K_2$ ($K_3K_4$), and $2C-N$ CMFMs $\gamma_{n>N}$ on edges $K_1K_4$ ($K_3K_2$) as shown in Fig.~\ref{fig1}(c). Different from the former case, there is only one path propagating from lead $1$ to lead $2$ (via $K_1K_2$), and only one path from lead $1$ back to itself (via $K_1K_4$), which contribute to $t-t_A$ and $r-r_A$, respectively. Accordingly, the $\lfloor N/2\rfloor$ CEMs $a_{j\le\lfloor N/2\rfloor}$ experience the SO($N$) transformation $Q$ on edge $K_1K_2$ and contribute to $t-t_A$, while the $\lfloor (2C-N)/2\rfloor$ CEMs $a_{j>\lfloor (N+1)/2\rfloor}$ participate in the SO($2C-N$) transformation $Q'$ on edge $K_1K_4$ and contribute to $r-r_A$.

Similarly, in the weak disorder regime, one arrives at the expression of $t(\overline{G})-t_A(\overline{G})$ and $r(\overline{G'})-r_A(\overline{G'})$ similar to Eq.~(\ref{ttaG}), where $\overline{G}$ and $\overline{G'}$ are the diagonalization matrices for $Q$ and $Q'$, respectively, which depends on sample and physical condition. By assuming $\overline{G}$ and $\overline{G'}$ distribute uniformly in coset spaces SO$(N)/$SO$(2)^{\lfloor N/2\rfloor}$ and SO$(2C-N)/$SO$(2)^{\lfloor (2C-N)/2\rfloor}$, respectively, one can calculate the conductance distributions. There are a few examples where the distribution $p(\sigma_{12})$ can be obtained analytically \cite{supple}, some of which are plotted in Fig.~\ref{fig2}(b). In particular, for $C=N=1$, the conductances are quantized at $\sigma_{12}=1/2$ and $\sigma_{13}=1$, in agreement with the previous results \cite{chung2011,wang2015c,he2017}.

Furthermore, the mean value $\overline{\sigma_{12}}=[C-\overline{(r-r_A)}+\overline{(t-t_A)}]/2$ for generic $0\le N<2C$ among various samples and physical conditions is derived as~\cite{supple}
\begin{equation}\label{conductance2}
\overline{\sigma_{12}}=
\begin{cases}
&\ \ \frac{C}{2}-\frac{1}{4N}+\frac{1}{4(2C-N)}\ ,\qquad (N\ \text{odd})\\
&\frac{C}{2}+\frac{1}{4(N-1)}-\frac{1}{4(2C-N-1)}\ .\ (N\ \text{even})\\
\end{cases}
\end{equation}

(c) \emph{$N\le0$ case.} There are $2C-N=2C+|N|$ CMFMs on edges $K_1K_4$ ($K_3K_2$), and $|N|$ backward propagating CMFMs on edges $K_1K_2$ ($K_3K_4$) as shown in Fig.~\ref{fig1}(d). Such a configuration is equivalent to a $\pi/2$ rotation of the $N>2C$ junction along with the interchanges $N\leftrightarrow 2C-N$ and $t,t_A\leftrightarrow r,r_A$. Therefore, the results of conductances in $N\ge2C$ case can be directly applied to the current case. Specifically, $t-t_A=0$, which indicates $\sigma_{13}\equiv2\sigma_{12}$. In the weak disorder regime, for $C=1$, we show the distribution of $\sigma_{12}\in[0,1/2]$ is \cite{supple}
\begin{equation}
p(\sigma_{12})=
\begin{cases}
&\delta(\sigma_{12})\ ,\qquad\qquad\qquad\quad (N=0)\\
& \left(\frac{1}{2}-\sigma_{12}\right)^{\alpha_N}{\sigma_{12}}^{\beta_N}\ ,\qquad (N<0)
\end{cases}
\end{equation}
where $\alpha_N=\frac{\lfloor -N/2\rfloor }{2}-\frac{1}{2}$ and $\beta_N=\frac{(2-N)(1-N)}{4}-\frac{\lfloor -N/2\rfloor }{2}-\frac{3}{2}$. The mean value of $\sigma_{12}$ for generic $C$ and $N\le0$ is
\begin{equation}\label{conductance3}
\overline{\sigma_{12}}=\frac{C}{2}\left[1-\frac{2\lfloor C-N/2\rfloor}{(2C-N)(2C-N-1)}\right]\ .
\end{equation}

We now turn to the strong disorder regime $|\delta L(x)|\gg |\overline{L}|$. In this limit, $\overline{G}$ is not well defined, and the SO($N$) transformation $Q$ should be regarded as fully random. The coefficient $t-t_A$ is then given by Eq. (\ref{tta}) averaged over all $Q$ under the Haar measure of group SO($N$), which is sample and physical condition \emph{independent}. Therefore, in (a) $N\geq2C$ case, $t-t_A=1$ for $N=2$, and $t-t_A=0$ for $N>2$~\cite{supple}. The conductances are quantized at $\sigma_{12}=1$, $\sigma_{13}=0$ for $N=2C=2$, and $\sigma_{12}=C/2$, $\sigma_{13}=C$ for $N>2$ and $N\ge2C$. In (b) $0\leq N<2C$ case, $t-t_A=1$ ($r-r_A=1$) if $N=2$ ($2C-N=2$), while $t-t_A=0$ ($r-r_A=0$) otherwise. Accordingly, the conductances are quantized at $\sigma_{12}=(C-\delta_{2C-N,2}+\delta_{N,2})/2$ and $\sigma_{13}=C-\delta_{2C-N,2}-\frac{\delta_{N,2}}{C-\delta_{2C-N,2}}$. Finally, in (c) $N<0$ case, the conductances are always quantized at $\sigma_{13}=2\sigma_{12}=C$.

%\emph{SO(N) vs. O(N)}.
%An important reason for the above unusual quantized conductances is that $N$ CMFMs on the TSC edge are undergoing a random SO($N$) instead of O($N$) transformation. In principle, if a CMFM $\gamma_n$ encounters a $\pi$-flux vortex that enters into (leaves) the TSC bulk across the edge, it will flip sign and may yield an O($N$) transformation outside the SO($N$) group. However, the vortices in the TSC bulk are gapped excitations, so such vortex creation/annihilation processes are generically exponentially suppressed, and only SO($N$) unitary transformations of CMFMs are allowed at zero temperature.

%\emph{Effects of the $\delta v$ term.}
Finally, we briefly discuss the effect of the irrelevant $\delta v$ term in Eq.~(\ref{Sv}), which is ignored in all the above three cases. In the above results, one may have noted that the junction conductance for $N=2$, $C=1$ is $\sigma_{12}=1$, which is the same as that of a single QHI. Namely, the $N=2$ CMFMs behave as a charge conserving CEM. This is generically not true, since an electron may turn into a hole and inject a Cooper pair into the TSC bulk, leading to a $\sigma_{12}<1$ \cite{lian2016a}. This discrepancy is because of our ignorance of the $\delta v$ term, which is proportional to the $p$-wave pairing (the only allowed pairing for $N=2$) on the TSC edge. Effectively, the presence of the $\delta v$ term makes the transformation matrix $Q$ effectively non-orthogonal (and nonunitary), and invalidates our conclusion  \cite{supple}.
%The reason that $Q$ is nonunitary is that $\delta v$ ($p$-wave pairing) leads to a probability of injecting a Cooper pair into the 2D TSC bulk, which is a process beyond the Hilbert space of CMFMs $\gamma_1$ and $\gamma_2$.
However, since the $\delta v$ term is irrelevant in the presence of disorders, its contribution will be suppressed to zero at large scales, and thus our conclusions in the above are valid for large enough systems.

\ \\
\emph{Thermal transport.} The chiral TSC with BdG Chern number $N$ exhibits a quantized thermal Hall conductance $\kappa_{xy}/T=N\kappa_0$ with $\kappa_0=(\pi^2/6)(k_B^2/h)$, where $k_B$ is the Boltzmann constant and $T$ is temperature. Interestingly, the random mixing of the CMFMs on the TSC edges will lead to quantized thermal resistance in the QHI-TSC-QHI junction.
It is similar to the quantized electric resistance in the $\nu$-$\nu'$-$\nu$ integer QH junction from the random mixing of the CEMs~\cite{kim2007}, where $\nu$ and $\nu'$ are filling fractions. For all the cases in Fig.~\ref{fig1}, when a heat current is applied between leads 1 and 2, the thermal resistances are given by $R^Q_{12,46}=R^Q_{12,57}=1/(2C)$ and
\begin{equation}
R^Q_{12,45}=R^Q_{12,67}=\frac{|2C-\text{sgn}(N)N|}{2C|N|}
\end{equation}
in units of $1/(\kappa_0T)$. Here we have assumed the full thermal equilibration among the copropagating CMFMs, which is appropriate when the sample size is large enough. If the thermal equilibration is not achieved, $R^Q_{12,45}$ and $R^Q_{12,67}$ will deviate from the above quantized value. Besides, phonons and magnons also contribute to the thermal resistance, but their contributions can be identified and subtracted by examining the temperature dependence~\cite{banerjee2017}.

\emph{Discussions.} We have shown the conductance of generic QHI-TSC-QHI junctions obey certain distributions in the weak disorder regime depending on $C$ and $N$, and tend to a quantized value in the strong disorder regime. For intermediate disorder strengths, we expect the conductance distributions to gradually evolve from the weak disorder limit to the strong disorder limit. In reality, the disorders are usually dilute, so are close to the weak disorder limit. In experiments, the conductance distribution can be measured by using different samples and varying physical conditions (gate voltage, in-plane and out-of-plane magnetic fields, etc.) within the phase space of the TSC. This will enable a comparison between our theory and the experiments.

If the TSC edges exhibit slow dynamical fluctuations (e.g., environmental fluctuation) compared to the propagation time of CMFMs on the edge $d/\overline{v}$, different electrons in the current will experience different physical conditions. Ideally, the time averaged $\sigma_{12}$ will then be given by Eqs. (\ref{conductance1}), (\ref{conductance2}) and (\ref{conductance3}), while the fluctuations of $\sigma_{12}$ obey the distribution $p(\sigma_{12})$.

In general, the TSC with an even $N=2C'$ can be obtained from the QHI with Chern number $C'$ under $s$-wave superconducting proximity, while the TSC with odd BdG Chern number $N=2C'-1$ can be obtained by pushing the $N=2C'$ TSC towards the QHI plateau transition point~\cite{qi2010b,chung2011,wang2015c}. In practice, one can use the heterostructure of topological nontrivial $s$-wave SC~\cite{zhang2018} and quantum anomalous Hall insulator (QAHI) with Chern number $C'$ \cite{jiang2018} to realize TSC with $N=2C'\pm1$, where keeps the disorders weak \cite{wang2018}. Furthermore, by creating a flipped magnetic domain in the QAHI and adding superconducting proximity on top of it, one can realize the junction with $N<0$ as shown in Fig.~\ref{fig1}(d).
At last, we note that although a QHI with Chern number $C'$ is topologically equivalent to a TSC with $N=2C'$, the transport of a charge conserving $C$-$C'$-$C$ QHI junction is quite different from that of the QHI-TSC-QHI junction. This is because an electron on TSC edges has a nonzero probability of becoming a hole due to the random rotation of CMFM basis, which is forbidden in the QHI junction.

\begin{acknowledgments}
\emph{Acknowledgments.}
B.L. is supported by the Princeton Center for Theoretical Science at Princeton University. J.W. is supported by the Natural Science Foundation of China through Grant No.~11774065; the National Key Research Program of China under Grant No.~2016YFA0300703; the Natural Science Foundation of Shanghai under Grant No.~17ZR1442500; the National Thousand-Young-Talents Program; the Open Research Fund Program of the State Key Laboratory of Low-Dimensional Quantum Physics, through Contract No.~KF201606; and by Fudan University Initiative Scientific Research Program.
\end{acknowledgments}

\newpage
\begin{widetext}
\section{Supplementary Material}

\section{Generic Multi-terminal Landauer-B\"uttiker formula}

The resistances measured from the other leads can be derived by generalizing the Landauer-B\"uttiker formula to the multi-lead case. In the configuration of main text Fig. 3, the multi-lead formula is
\begin{equation}
\begin{split}
&I_1=\frac{e^2}{h}(V_1-V_6)\ ,\quad I_6=\frac{e^2}{h}[(V_6-V_{sc})+(t-t_A)(V_{sc}-V_7)+(r-r_A)(V_{sc}-V_4)]\ ,\quad I_7=\frac{e^2}{h}(V_7-V_2)\ ,\\
&I_2=\frac{e^2}{h}(V_2-V_5)\ ,\quad I_5=\frac{e^2}{h}[(V_5-V_{sc})+(t-t_A)(V_{sc}-V_4)+(r-r_A)(V_{sc}-V_7)]\ ,\quad I_4=\frac{e^2}{h}(V_4-V_1)\ ,\\
&I_3=-(I_1+I_2+I_4+I_5+I_6+I_7)\ ,\quad V_3=V_{sc}\ ,
\end{split}
\end{equation}
where $r$, $r_A$, $t$ and $t_A$ are the coefficients we defined above. Solving the equations with respect to applied currents yield the quantized values of resistances measured on various leads.

\section{The $\delta v$ term}
In the main text Eq. (3), we show the action can be rewritten as an SO($N$) symmetric term and a $\delta v$ term $i\widetilde{\gamma}^TQ^T\delta v\partial_x(Q\widetilde{\gamma})$. Here we first show the $\delta v$ term is irrelevant.

In the weak disorder limit, we show the matrix $Q$ can be approximately written as $Q=\overline{G}\Lambda(x)\overline{G}^T$, where if we take odd $N$ as an example $\Lambda(x)=\mbox{diag}(e^{i\phi_1(x)\sigma_y},\cdots,1)$. In particular, the angles $\phi_l(x)$ $(1\le l\le\lfloor N/2\rfloor)$ can be regarded as fully random, namely, $\overline{e^{i\phi_1(x)-i\phi_1(x')}}=\delta(x-x')$. The $\delta v$ term can then be written as
\begin{equation}
H_{\delta v}
%=-i\widetilde{\gamma}^TQ^T\delta vQ\partial_x\widetilde{\gamma}- i \widetilde{\gamma}^T(Q^T\delta v\partial_xQ)\widetilde{\gamma}
=-i\widetilde{\gamma}^TJ_1(x)\partial_x\widetilde{\gamma}- i \widetilde{\gamma}^TJ_2(x)\widetilde{\gamma}\ ,
\end{equation}
where $J_1(x)=\overline{G}\Lambda^T(x)\delta v_G\Lambda(x)\overline{G}^T$ and $J_2(x)=\overline{G}\Lambda^T(x)\delta v_G\partial_x\Lambda(x)\overline{G}^T$. Accordingly, one find the correlation functions
\begin{equation}
\begin{split}
&\overline{\mbox{Tr}[J_1(x)J_1(x')]}=\overline{\mbox{Tr}(\delta v_G\Lambda(x-x')\delta v_G\Lambda(x'-x))}\sim W_1\delta(x-x')\ ,\\
&\overline{\mbox{Tr}[J_2(x)J_2(x')]}=\overline{\mbox{Tr}(\delta v_G\partial_x\Lambda(x-x')\delta v_G\partial_{x'}\Lambda(x'-x))}\sim W_2\partial_{x'}\partial_{x}\delta(x-x')\ ,
\end{split}
\end{equation}
where $\Lambda(x'-x)=\mbox{diag}(e^{i(\phi_1(x)-\phi_1(x'))\sigma_y},\cdots,1)$, and we assumed $\delta v_G$ does not commute with $\Lambda(x'-x)$, which is usually true. Therefore, one finds the scaling dimensions of $W_1$ and $W_2$ are $d_{W_1}=d_{W_2}=-1<0$, which indicates the $\delta v$ term is irrelevant. Such an analysis can also be done for the strong disorder limit, and the conclusion remains valid.

Although the $\delta v$ term is irrelevant, we can still discuss the role it plays. We shall take the minimal example of $N=2$, where there are two CMFMs $\gamma_1$ and $\gamma_2$ consisting of an electron charge basis $a_1,a_1^\dag=(\gamma_1\pm i\gamma_2)/\sqrt{2}$. The CMFM action rewritten in the charge basis is
\begin{equation}
S=\int dtdx ia_1^\dag(\partial_t-\overline{v} \partial_x)a_1+\left(\Delta_p a_1^\dag\partial_x a_1^\dag +h.c.\right)-\mu(x)a_1^\dag a_1\ ,
\end{equation}
where $\mu(x)=2L_{12}(x)$ is given by the off diagonal element of $L(x)$, the mean velocity $\overline{v}=(v_1+v_2)/2$, while the $p$-wave pairing amplitude $\Delta_p=(v_1-v_2)/4=\delta v/2$ results from the $\delta v$ term. Therefore, if $\delta v$ is ignored, the action will behave as a charge conserved CEM with no superconductivity. This is in agreement with our main text Eq. (4): if one take $C=1$ and $N=2$, one finds $t=1$ and $t_A=0$, which is no different from a charge conserved QHI junction. This is because the SO($2$) rotation cannot rotate an electron into a hole. The ignorance of $\delta v$ is legitimate in large samples since it is irrelevant.

However, if we consider a system with no disorder, the $\delta v$ term becomes marginal and cannot be ignored. As is shown in Ref. \cite{lian2016a}, for an $N=2$ junction without disorder, the Andreev probability $t_A$ is generically nonzero, since a chiral electron on the edge can turn into a hole with a Cooper pair injected into the TSC bulk. To see this explicitly, let us consider a system with no disorder so that $\mu$ is constant, and solve the Euler-Lagrange equation at zero energy:
\begin{equation}
\left(\begin{array}{cc}
-iv_1\partial_x & -i\mu\\
i\mu & -iv_2\partial_x\\
\end{array}\right)
\left(\begin{array}{c}
\gamma_1 \\ \gamma_2
\end{array}\right)=0\ ,
\end{equation}
Assume the incident state is an electron state $\gamma(0)=(1,i)^T/\sqrt{2}$ which enters at $x=0$. For small $\delta v=(v_1-v_2)/2$, solving the equation yields a solution
\begin{equation}
\begin{split}
&\gamma(x)\approx\frac{1}{\sqrt{2}}\left(\begin{array}{c}1-\frac{k_0\delta v}{\mu}\\ i+i\frac{k_0\delta v}{\mu}\end{array}\right)e^{ik_0x}+\frac{1}{\sqrt{2}}\frac{k_0\delta v}{\mu}\left(\begin{array}{c}1+\frac{k_0\delta v}{\mu}\\ -i+ i\frac{k_0\delta v}{\mu}\end{array}\right)e^{-ik_0x}\\
\approx &\left(\begin{array}{cc}\cos k_0x & (1-2\frac{k_0\delta v}{\mu})\sin k_0x\\ -(1+2\frac{k_0\delta v}{\mu})\sin k_0x & \cos k_0x\end{array}\right)\left(\begin{array}{c}1\\ i\end{array}\right)/\sqrt{2}= Q'(x)\gamma(0)\ ,
\end{split}
\end{equation}
where $k_0=\mu/\sqrt{v_1v_2}$. In particular, one finds the transformation matrix $Q'(x)$ is no longer an orthogonal matrix (it is even not a unitary matrix). If we keep using it, we find a nonzero Andreev transmission
\[
t_A\approx\left|(\frac{1}{\sqrt{2}},\frac{i}{\sqrt{2}})Q'(x)\left(\begin{array}{c}\frac{1}{\sqrt{2}}\\ \frac{i}{\sqrt{2}}\end{array}\right)\right|^2\approx \left(\frac{4k_0\delta v}{\mu}\right)^2\sin^2 k_0x\ .
\]

The fact that $Q'(x)$ becomes nonunitary with $\delta v\neq 0$ can be understood as follows. We have shown that $\delta v$ for $N=2$ is nothing but the $p$-wave pairing amplitude on the TSC edge. Therefore, $\delta v$ induces a probability for an electron to turn into a hole, and inject a Cooper pair into the TSC bulk at the same time. However, the Cooper pair injected into the 2D TSC bulk is beyond the Hilbert space of the $1+1$ dimensional TSC edge. Therefore, the transformation in the Hilbert space of $\gamma_1$ and $\gamma_2$ becomes effectively nonunitary (thus non-orthogonal). This conclusion generically applies for $N>2$ as well, namely, the $\delta v$ term yields a nonunitary (non-orthogonal) contribution to the $N\times N$ transformation matrix $Q(x)$.

In the disordered case, as we showed the $\delta v$ term is irrelevant, so we do not need to worry about this nonunitary problem.

\section{Derivation of the junction conductance distributions for $N\ge2C$}

We begin our calculation from the main text Eq. (4). From the definition, it is straightforward to show that $|u_j^\dag Qu_i|^2-|u_j^T Qu_i|^2=\det Q_2^{(2j-1,2j|2i-1,2i)}$, where
\[
Q_2^{(i,j|k,l)}=\left(\begin{array}{cc}
Q_{ik}& Q_{il}\\
Q_{jk} & Q_{jl} \\
\end{array}\right)
\]
is the $2\times2$ minor matrix of the $i$-th, $j$-th rows and $k$-th, $l$-th columns of $Q$ (assume $i<j$ and $k<l$ hereafter), and $Q_{ik}$ is the matrix element. Therefore, we can rewrite the main text Eq. (4) as
\begin{equation}
t-t_A=\sum_{i,j=1}^{C}\det Q_2^{(2j-1,2j|2i-1,2i)}\ .
\end{equation}

\subsection{Generic formula for weak disorder}
In the weak disorder case, we approximately have $Q=\overline{G}\Lambda\overline{G}^T$, where  $\Lambda=\text{diag}(e^{i\phi_1\sigma_y},\cdots,e^{i\phi_{N/2}\sigma_y})$ for even $N$, and $\Lambda=\text{diag}(e^{i\phi_1\sigma_y},\cdots,e^{i\phi_{(N-1)/2}\sigma_y},1)$ for odd $N$. We also denote $\overline{G}=(\bm{g}_1,\cdots,\bm{g}_N)$, where the vectors $\bm{g}_l=(g^1_l,\cdots,g^N_l)^T$ $(1\le l\le N)$ form a new orthonormal Majorana basis. To obtain the decoherent $t-t_A$, we need to calculate $\det Q_2^{(2j-1,2j|2i-1,2i)}$ averaged over all angles $\phi_l$. To further simplify the calculation, we can fully diagonalize matrix $\Lambda$ using $\sigma_y=(\frac{1+i\sigma_x}{\sqrt{2}})\sigma_z(\frac{1-i\sigma_x}{\sqrt{2}})$, after which we have for odd $N$
\begin{equation}
Q=\left(\frac{\bm{g}_1+i\bm{g}_2}{\sqrt{2}},\frac{i\bm{g}_1+\bm{g}_2}{\sqrt{2}},\cdots,\bm{g}_N\right) \text{diag}(e^{i\phi_1},e^{-i\phi_1},\cdots,e^{i\phi_l},e^{-i\phi_l},\cdots,1) \left(\frac{\bm{g}_1-i\bm{g}_2}{\sqrt{2}},\frac{-i\bm{g}_1+\bm{g}_2}{\sqrt{2}},\cdots,\bm{g}_N\right)^T\ ,
\end{equation}
and similarly for even $N$. Hereafter we shall take odd $N$ as an example, while the even $N$ case can be obtained by simply deleting the last component from the odd $N$ case. A direct calculation gives the expression of the matrix element
\begin{equation}
\begin{split}
&Q_{jk}=\sum_{l=1}^{\lfloor N/2\rfloor}\left[e^{i\phi_l}\left(\frac{g_{2l-1}^j+ig_{2l}^j}{\sqrt{2}}\right) \left(\frac{g_{2l-1}^k-ig_{2l}^k}{\sqrt{2}}\right)+ e^{-i\phi_l}\left(\frac{g_{2l-1}^j-ig_{2l}^j}{\sqrt{2}}\right) \left(\frac{g_{2l-1}^k+ig_{2l}^k}{\sqrt{2}}\right) \right]+g_N^jg_N^k\ .
\end{split}
\end{equation}
The determinant $\det Q_2^{(j,j'|k,k')}=Q_{jk}Q_{j'k'}-Q_{jk'}Q_{j'k}$ generically contain many terms. When averaged over all phases $\phi_l$, the only nonzero terms are those contain the product of conjugate phase factors $e^{i\phi_l}$ and $e^{-i\phi_l}$, or the $g_N$ terms which have no phase factors. Explicitly, we have
\begin{equation}
\begin{split}
&\overline{\det Q_2^{(j,j'|k,k')}}=\frac{1}{4}\sum_{l=1}^{\lfloor N/2\rfloor} \Big\{\Big[ \left(g_{2l-1}^j+ig_{2l}^j\right) \left(g_{2l-1}^k-ig_{2l}^k\right) \left(g_{2l-1}^{j'}-ig_{2l}^{j'}\right) \left(g_{2l-1}^{k'}+ig_{2l}^{k'}\right)\\
&\quad-\left(g_{2l-1}^j+ig_{2l}^j\right) \left(g_{2l-1}^{k'}-ig_{2l}^{k'}\right) \left(g_{2l-1}^{j'}-ig_{2l}^{j'}\right) \left(g_{2l-1}^{k}+ig_{2l}^{k}\right)\Big]+h.c.\Big\} +g_N^jg_N^kg_N^{j'}g_N^{k'}-g_N^jg_N^{k'}g_N^{j'}g_N^k\\
&=\sum_{l=1}^{\lfloor N/2\rfloor} \left(g_{2l-1}^{j}g_{2l}^{j'}-g_{2l-1}^{j'}g_{2l}^{j}\right) \left(g_{2l-1}^{k}g_{2l}^{k'}-g_{2l-1}^{k'}g_{2l}^{k}\right)\\
&=\sum_{l=1}^{\lfloor N/2\rfloor} A_{2l-1,2l}^{jj'} A_{2l-1,2l}^{kk'}\ ,
\end{split}
\end{equation}
where we have defined $A_{lk}^{ij}=g_{l}^{i}g_{k}^{j}-g_{l}^{j}g_{k}^{i}$, which is mathematically the projection of unit area in the $(\bm{g}_l,\bm{g}_{k})$ plane onto the original basis $(\gamma_i,\gamma_{j})$ plane. This indicates they satisfy the following conditions
\begin{equation}\label{Cons1}
A_{lk}^{ij}=-A_{lk}^{ji}=-A_{kl}^{ij}\ ,\qquad\qquad \sum_{1=l<k}^N(A_{lk}^{ij})^2=1 \quad(i\neq j)\ ,\qquad\qquad \sum_{1=i<j}^N(A_{lk}^{ij})^2=1 \quad(l\neq k)\ .
\end{equation}
With the above result, we find
\begin{equation}
t(\overline{G})-t_A(\overline{G})=\sum_{l=1}^{\lfloor N/2\rfloor}\left(\sum_{j=1}^C A_{2l-1,2l}^{2j-1,2j}\right)^2\ ,
\end{equation}
namely, the main text Eq. (5). We note this formula is also valid for even $N$, since the last component $g_N$ in the odd $N$ case does not contribute this formula. For $N\ge2C$, one then has $\sigma_{12}=(C+t-t_A)/2$, and $\sigma_{13}=C-\frac{(t-t_A)^2}{C}$.

To proceed to calculate the conductance distribution, we need to understand better the coset space SO$(N)/$SO$(2)^{\lfloor N/2\rfloor}$ the matrix $\overline{G}$ lives in. In the $\gamma_n$ basis, the generators of the SO($N$) group are denoted by $T_{ab}=-T_{ba}$ ($1\le a<b\le N$), which are $N\times N$ matrices with the element in the $a$-th ($b$-th) row, $b$-th ($a$-th) column being $-i$ ($i$), and all the other elements zero. They satisfy the commutation relation
\begin{equation}
[T_{ab},T_{cd}]=-i(\delta_{bc}T_{ad}+\delta_{ad}T_{bc}-\delta_{bd}T_{ac}-\delta_{ac}T_{bd})\ .
\end{equation}
In the linear space of the $N(N-1)/2$ generators $T_{ab}$, one can maximally select out $\lfloor N/2\rfloor$ generators $H_l$ ($1\le l\le\lfloor N/2\rfloor$) which mutually commute, which is known as the Cartan subalgebra. For the matrix $\overline{G}$ here, the $\lfloor N/2\rfloor$ generators $H_l$ are given by $N\times N$ matrices
\begin{equation}
H_l=(\bm{g}_{2l-1},\bm{g}_{2l})\sigma_y(\bm{g}_{2l-1},\bm{g}_{2l})^T=\sum_{1=a<b}^N A_{2l-1,2l}^{ab}T_{ab}\ ,
\end{equation}
which generates the rotation in the $(\bm{g}_{2l-1},\bm{g}_{2l})$ plane. From the orthogonality of $\bm{g}_l$, one can easily show $[H_l,H_{k}]=0$, and accordingly
\begin{equation}
[H_l,H_{k}]=-i\sum_{a,b,c=1}^N A_{2l-1,2l}^{ab}A_{2k-1,2k}^{bc}T_{ac}=0\ ,
\end{equation}
which implies another constraint for $A_{2l-1,2l}^{ab}$
\begin{equation}\label{Cons2}
\sum_{b=1}^N (A_{2l-1,2l}^{ab}A_{2k-1,2k}^{bc}-A_{2l-1,2l}^{cb}A_{2k-1,2k}^{ba})=0\ .
\end{equation}
The coset space SO$(N)/$SO$(2)^{\lfloor N/2\rfloor}$ is then parameterized by $H_l$, which can be expressed in terms of coefficients $A_{2l-1,2l}^{ab}$. However, the coefficients $A_{2l-1,2l}^{ab}$ are not independent, but subject to the constraints of the supplementary material Eqs. (\ref{Cons1}) and (\ref{Cons2}) and thus are highly correlated. This makes the measure of the coset space and thus the calculation of the probability distribution of $t-t_A$ extremely complicated. We shall not try to solve for the coset space measure here, instead we study a simple case of Chern number $C=1$ (and $N\ge2C$) in the following.

\subsection{Conductance Distribution for $N\ge2C$ with $C=1$}

In the case $C=1$ (and $N\ge2C$), the coefficient $t-t_A$ is simply given by
\begin{equation}
t(\overline{G})-t_A(\overline{G})=\sum_{l=1}^{\lfloor N/2\rfloor}\left(A_{2l-1,2l}^{1,2}\right)^2\ ,
\end{equation}
while we know $\sum_{1=l<k}^N\left(A_{lk}^{1,2}\right)^2=1$, so the $M=N(N-1)/2$ coefficients $A_{lk}^{1,2}$ live on an $M-1$ dimensional unit sphere $S^{M-1}$. If we assume the distribution of $A_{lk}^{1,2}$ is uniform on $S^M$, we can easily calculate the distribution of $t(\overline{G})-t_A(\overline{G})$. This can be most easily done by assuming $\sum_{l=1}^{\lfloor N/2\rfloor}\left(A_{2l-1,2l}^{1,2}\right)^2=\cos^2\theta$, where $0\le\theta\le\pi/2$. This means (for $N\ge3$) the $\lfloor N/2\rfloor$ coefficients $A_{2l-1,2l}^{1,2}$ lives on a sphere $S^{p}$ of dimension $p=\lfloor N/2\rfloor-1$ with radius $r_1=\cos\theta$, while the rest coefficients $A_{lk}^{1,2}$ lives on a sphere $S^{q}$ of dimension $q=M-\lfloor N/2\rfloor-1$ with radius $r_2=\sin\theta$. The unit sphere $S^{M-1}$ can then be parameterized by $\theta$ and the coordinates of sphere $S^p$ and of sphere $S^q$. In particular, the measure $d\mu(\theta)$ for parameter $\theta$ is proportional to the area of sphere $S^p$ and area of sphere $S^q$, so the unnormalized measure has the form
\begin{equation}\label{mu}
d\mu(\theta)=r_1^pr_2^qd\theta=\cos^p\theta\sin^q\theta d\theta\qquad (N>2)\ .
\end{equation}
For instance, in the familiar SO($3$) case $N=3$, one has $p=0$, $q=1$, and thus $d\mu(\theta)=\sin\theta d\theta$. While a special case is when $N=2$, one has $M=\lfloor N/2\rfloor=1$, thus $t(\overline{G})-t_A(\overline{G})\equiv1$. In this case, the above formula is not valid, instead one has $d\mu(\theta)=\delta(\theta)d\theta$. The (unnormalized) probability distribution of $\sigma_{12}=(1+t-t_A)/2= (1+\cos^2\theta)/2$ for $C=1$ can then be calculated:
\begin{equation}
p(\sigma_{12})=\frac{d\mu(\theta)}{d\sigma_{12}}=\frac{d\mu(\theta)}{\sin\theta\cos\theta d\theta}
\begin{cases}
&\propto\delta(\theta)\propto\delta(\sigma_{12}-1)\qquad\qquad\qquad\qquad\qquad\qquad\quad (N=2) \ ,\\
&=\cos^{p-1}\theta\sin^{q-1}\theta\propto \left(\sigma_{12}-\frac{1}{2}\right)^{\alpha_N}\left(1-\sigma_{12}\right)^{\beta_N}\qquad (N>2)\ ,
\end{cases}
\end{equation}
where the power indices $\alpha_N=\frac{p-1}{2}=\frac{\lfloor N/2\rfloor }{2}-1$ and $\beta_N=\frac{q-1}{2} =\frac{N(N-1)}{4}-\frac{\lfloor N/2\rfloor }{2}-1$. After normalization, the probability distribution of $\sigma_{12}$ for $N>2$ reads
\[
p(\sigma_{12})=\frac{2^{\alpha_N+\beta_N+1}\Gamma(\alpha_N+\beta_N+2)}{ \Gamma(\alpha_N+1)\Gamma(\beta_N+1)} \left(\sigma_{12}-\frac{1}{2}\right)^{\alpha_N}\left(1-\sigma_{12}\right)^{\beta_N}\ .
\]
Similarly, one can derive the distribution of $\sigma_{13}=1-(t-t_A)^2=1-\cos^4\theta$ for $C=1$ to be
\begin{equation}
p(\sigma_{13})=\frac{d\mu(\theta)}{d\sigma_{13}}=\frac{d\mu(\theta)}{4\sin\theta\cos^3\theta d\theta}
\begin{cases}
&\propto\delta(\sigma_{13})\qquad\qquad\qquad\qquad\qquad\qquad\qquad\qquad\qquad (N=2) \ ,\\
&=\cos^{p-3}\theta\sin^{q-1}\theta\propto \left(1-\sigma_{13}\right)^{\frac{\alpha_N-1}{2}}\left(1-\sqrt{1-\sigma_{13}}\right)^{\beta_N}\quad (N>2)\ .
\end{cases}
\end{equation}

\subsection{Mean value of conductances for generic $N>2C$}
For generic Chern number $C$, the distribution of $\sigma_{12}$ can hardly be calculated analytically, thus we shall not try to do so. However, the mean value of the junction conductance $\sigma_{12}$ can be easily calculated for any $N$ and $C$. This is done by noting that the normalization condition in Eq. (\ref{Cons1}) implies the mean value
\begin{equation}\label{meanA}
\overline{\left(A_{kl}^{ij}\right)^2}=\frac{2}{N(N-1)}\ ,\qquad \overline{A_{kl}^{ij}A_{k'l'}^{i'j'}}=0\qquad (\{i,j,k,l\}\neq\{i',j',k',l'\})\ ,
\end{equation}
since all $A_{kl}^{ij}$ should obey the same distribution, and two different coefficients $A_{kl}^{ij}$ and $A_{k'l'}^{i'j'}$ can be independently positive or negative. Therefore, the mean value of $\sigma_{12}$ (in units of $e^2/h$) for $N\ge 2C$ is given by
\begin{equation}
\overline{\sigma_{12}}=\frac{C+\overline{t-t_A}}{2}=\frac{C}{2}\left[1+\frac{2\lfloor N/2\rfloor}{N(N-1)}\right]\ .
\end{equation}
\subsection{Conductance in the strong disorder case}
In the strong disorder case $|\delta L(x)|\gg|\overline{L}|$, the SO($N$) matrix $Q$ can be consider as fully random under the decoherent assumption. In this case, $t-t_A$ can be considered as the average of the main text Eq. (4) over all SO($N$) matrices $Q$ upon the Haar measure of SO($N$), which is no longer sample dependent but is a unique value.

To find the value of $t-t_A$, we simply note that the Haar measure $d\mu_h$ of group SO($N$) satisfies $d\mu_h(Q)=d\mu_h(PQ)$ for any $P,Q\in$SO($N$). Consider now a $2\times2$ minor $\text{det}Q_2^{(ij|kl)}$ of matrix $Q$. For $N>2$, we can choose a matrix $P\in$SO($N$) as $P=e^{i\pi T_{jj'}}$ with $j'\neq i$. Multiplying $P$ on the left of $Q$ then flips sign of the $j$-th and $j'$-th rows of $Q$, and thus flips sign of $\text{det}Q_2^{(ij|kl)}$. Therefore, one finds the mean value of the minor is
\begin{equation}
\overline{\text{det}Q_2^{(ij|kl)}}=\int d\mu_h(Q)\text{det}Q_2^{(ij|kl)}= \int d\mu_h(PQ)\text{det}(PQ)_2^{(ij|kl)}=-\int d\mu_h(Q)\text{det}Q_2^{(ij|kl)}=0\ .
\end{equation}
This implies $t-t_A\equiv 0$ under the average over the Haar measure of SO($N$). Accordingly, the conductances in the strong disorder limit are quantized to
\begin{equation}
\sigma_{12}=\frac{C}{2}\ ,\qquad\qquad\sigma_{13}=C \ ,
\end{equation}
for any $N>2$ and $N\ge2C$. This simply means the electron always have $1/2$ probability to become an electron or hole.

The above calculation can be directly generalized to any $N$ and $C$ in the strong disorder limit, from which the quantized conductances can be summarized into
\begin{equation}\label{strong}
\sigma_{12}=(C-\delta_{2C-N,2}+\delta_{N,2})/2\ ,\qquad \sigma_{13}=C-\delta_{2C-N,2}-\frac{\delta_{N,2}}{C-\delta_{2C-N,2}}\ .
\end{equation}

\section{Conductance Distribution for $0<N<2C$}

The conductance in the $0<N<2C$ case is more complicated, since both $t-t_A$ and $r-r_A$ are nonzero, which are given by SO$(N)$ and SO$(2C-N)$ rotations of CMFMs on edges $K_1K_2$ and $K_1K_4$, respectively. The chiral fermion mode $a_j$ with $j\le \lfloor N/2\rfloor$ participates in the SO($N$) rotation $Q$ on edge $K_1K_2$ and contributes to $t-t_A$, while $a_j$ with $j\ge \lfloor (N+1)/2\rfloor$ participates in the SO($2C-N$) rotation $Q'$ on edge $K_1K_4$ and contributes to $r-r_A$.

In the weak disorder case, one has the approximations $Q=\overline{G}\Lambda \overline{G}^T$ and $Q'=\overline{G'}\Lambda'\overline{G'}^T$, where $\overline{G}=(\bm{g}_1,\cdots, \bm{g}_N)$ and $\overline{G'}=(\bm{g}_1',\cdots, \bm{g}_{2C-N}')$ are SO($N$) and SO($2C-N$) matrices, respectively. Following the same procedure we have done above, one finds
\begin{equation}
t(\overline{G})-t_A(\overline{G})=\sum_{l=1}^{\lfloor N/2\rfloor}\left(\sum_{j=1}^{\lfloor N/2\rfloor} A_{2l-1,2l}^{2j-1,2j}\right)^2\ ,\qquad
r(\overline{G'})-r_A(\overline{G'})=\sum_{l=1}^{\lfloor (2C-N)/2\rfloor}\left(\sum_{j=1}^{\lfloor (2C-N)/2\rfloor} {A'}_{2l-1,2l}^{2j-1,2j}\right)^2\ ,
\end{equation}
where $A_{lk}^{ij}$ and ${A'}_{lk}^{ij}$ are the projected unit area of matrices $\overline{G}$ and $\overline{G}'$, respectively. Accordingly, one can compute $\sigma_{12}$ and $\sigma_{13}$.

The distribution of conductances for generic $0<N<2C$ is difficult. However, similar to our previous discussion, we are able to calculate the distribution for $0<N<2C$ when both $\lfloor N/2\rfloor\le1$ and $\lfloor (2C-N)/2\rfloor\le1$. There are $5$ cases this condition can be satisfied when $C\le 3$, which we list as follows.

(1) $C=1$, $N=1$. This case is trivial, one has $t-t_A\equiv r-r_A\equiv0$, and quantized conductances $\sigma_{12}=1/2$ and $\sigma_{13}=1$.

(2) $C=2$, $N=3$. In this case $\lfloor N/2\rfloor=1$ and $\lfloor (2C-N)/2\rfloor=0$, thus $r-r_A=0$, while following Eq. (\ref{mu}) we have $t(\overline{G})-t_A(\overline{G})=\cos^2\theta$, with the measure of $\overline{G}$ given by $d\mu(\theta)=\sin\theta d\theta$. Accordingly, $\sigma_{12}=(2+\cos^2\theta)/2$ and $\sigma_{13}=2-\cos^4\theta/2$. Therefore, one find the distribution of conductances
\begin{equation}
p(\sigma_{12})=\frac{d\mu(\theta)}{d\sigma_{12}}\propto(\sigma_{12}-1)^{-1/2}\ ,\qquad p(\sigma_{13})=\frac{d\mu(\theta)}{d\sigma_{13}}\propto(2-\sigma_{13})^{-3/4}\ ,
\end{equation}
where $\sigma_{12}\in[1,3/2]$, and $\sigma_{13}\in[3/2,2]$.

(3) $C=2$, $N=2$. In this case $\lfloor N/2\rfloor=\lfloor (2C-N)/2\rfloor=1$, and thus we have $t(\overline{G})-t_A(\overline{G})=\cos^2\theta$ and $r(\overline{G'})-r_A(\overline{G'})=\cos^2\theta'$, where the measure is given by $d\mu(\theta,\theta')=\sin\theta\sin\theta' d\theta d\theta'$. This leads to $\sigma_{12}=(2-\cos^2\theta'+\cos^2\theta)/2$ and $\sigma_{13}=2-\cos^2\theta'-\cos^4\theta/(2-\cos^2\theta')$. The probability distribution of $\sigma_{12}\in[1/2,3/2]$ can be calculated to be
\begin{equation}
p(\sigma_{12})=\frac{\int_{\sigma_{12}}^{\sigma_{12}+d\sigma_{12}} d\mu}{d\sigma_{12}} = \frac{1}{2\sqrt{1-2|\sigma_{12}-1|}} +\ln\left(\frac{1+\sqrt{1-2|\sigma_{12}-1|}}{\sqrt{2|\sigma_{12}-1|}}\right) -\frac{1}{2}-\frac{|\sigma_{12}-1|}{1-2|\sigma_{12}-1|+\sqrt{1-2|\sigma_{12}-1|}}.
\end{equation}
The distribution of $\sigma_{13}$ will not be calculated explicitly, which is much more complicated.

(4) $C=2$, $N=1$. In this case $\lfloor N/2\rfloor=0$ and $\lfloor (2C-N)/2\rfloor=1$, and thus we have $t-t_A\equiv0$ and $r(\overline{G'})-r_A(\overline{G'})=\cos^2\theta'$, with the measure $d\mu(\theta') =\sin\theta'd\theta'$. The conductances are given by $\sigma_{12}=(2-\cos^2\theta')/2$ and $\sigma_{13}\equiv2$. The distribution of $\sigma_{12}\in[1/2,1]$ is given by
\begin{equation}
p(\sigma_{12})=\frac{d\mu(\theta')}{d\sigma_{12}}\propto(1-\sigma_{12})^{-1/2}\ .
\end{equation}

(5) $C=3$, $N=3$. In this case again $\lfloor N/2\rfloor=\lfloor (2C-N)/2\rfloor=1$, and we have $t(\overline{G})-t_A(\overline{G})=\cos^2\theta$ and $r(\overline{G'})-r_A(\overline{G'})=\cos^2\theta'$, where the measure is given by $d\mu(\theta,\theta')=\sin\theta\sin\theta' d\theta d\theta'$. The conductances are given by $\sigma_{12}=(3-\cos^2\theta'+\cos^2\theta)/2$ and $\sigma_{13}=3-\cos^2\theta'-\cos^4\theta/(3-\cos^2\theta')$. The distribution of $\sigma_{12}\in[1,2]$ can also be computed to be
\begin{equation}
p(\sigma_{12}) = \frac{1}{2\sqrt{1-|2\sigma_{12}-3|}} +\ln\left(\frac{1+\sqrt{1-|2\sigma_{12}-3|}}{\sqrt{|2\sigma_{12}-3|}}\right)-\frac{1}{2} -\frac{|\sigma_{12}-3/2|}{1-|2\sigma_{12}-3|+\sqrt{1-|2\sigma_{12}-3|}}.
\end{equation}

For other $C$ and $N$ satisfying $0<N<2C$, we cannot solve for the distribution, but again we can calculate the mean value of $\sigma_{12}$. Using supplementary Eq. (\ref{meanA}), we can easily find
\begin{equation}
\overline{t-t_A}=\frac{2\lfloor N/2\rfloor^2}{N(N-1)}\ ,\qquad \overline{r-r_A}=\frac{2\lfloor C-N/2\rfloor^2}{(2C-N)(2C-N-1)}\ ,
\end{equation}
and thus
\begin{equation}
\overline{\sigma_{12}}=\frac{\overline{C-r+r_A+t-t_A}}{2}=
\begin{cases}
&\ \ \frac{C}{2}-\frac{1}{4N}+\frac{1}{4(2C-N)}\ ,\qquad \quad(N\ \text{odd})\\
&\frac{C}{2}+\frac{1}{4(N-1)}-\frac{1}{4(2C-N-1)}\ .\quad(N\ \text{even})\\
\end{cases}
\end{equation}

In the strong disorder case, the conductances are as shown in supplementary Eq. (\ref{strong}).

\section{Conductance Distribution for $N\le0$}

The $N<0$ case can be viewed as the $N\ge 2C$ device rotated by $\pi/2$, and substituting $N\leftrightarrow 2C-N$, and $t,t_A\leftrightarrow r,r_A$. Accordingly, the rest derivations directly follows. In particular, we have $t-t_A\equiv0$, from which one can show $\sigma_{13}\equiv2\sigma_{12}$. Therefore, we shall only present the results for $\sigma_{12}$.

In the weak disorder case, again for $C=1$, the distribution of $\sigma_{12}$ can be derived, which takes the form
\begin{equation}
p(\sigma_{12})=
\begin{cases}
&\delta(\sigma_{12})\qquad\qquad\qquad\quad (N=0) \ ,\\
& \left(\frac{1}{2}-\sigma_{12}\right)^{\alpha_N}{\sigma_{12}}^{\beta_N}\qquad (N<0)\ ,
\end{cases}
\end{equation}
where $\sigma_{12}\in[0,1/2]$, and $\alpha_N=\frac{\lfloor (2-N)/2\rfloor }{2}-1$ and $\beta_N=\frac{(2-N)(2-N-1)}{4}-\frac{\lfloor (2-N)/2\rfloor }{2}-1$.

Besides, if $N=0$, the conductance $\sigma_{12}\equiv0$ for any $C$, which is obvious.

For more generic $C$ and $N<0$, the distribution is hard to calculate, but again the mean value of $r-r_A$ and thus $\sigma_{12}$ can be easily derived to be
\begin{equation}
\overline{r-r_A}=\frac{2C\lfloor (2C-N)/2\rfloor}{(2C-N)(2C-N-1)} \ ,
\end{equation}
and
\begin{equation}
\sigma_{12}=\frac{\overline{C-r+r_A}}{2}=\frac{C}{2}\left[1-\frac{2\lfloor (2C-N)/2\rfloor}{(2C-N)(2C-N-1)}\right]\ .
\end{equation}

In the strong disorder limit, one can show $t-t_A=r-r_A=0$, and $\sigma_{13}=2\sigma_{12}=C$.

\end{widetext}

\bibliography{ntsc_ref}

\begin{thebibliography}{31}%
\makeatletter
\providecommand \@ifxundefined [1]{%
 \@ifx{#1\undefined}
}%
\providecommand \@ifnum [1]{%
 \ifnum #1\expandafter \@firstoftwo
 \else \expandafter \@secondoftwo
 \fi
}%
\providecommand \@ifx [1]{%
 \ifx #1\expandafter \@firstoftwo
 \else \expandafter \@secondoftwo
 \fi
}%
\providecommand \natexlab [1]{#1}%
\providecommand \enquote  [1]{``#1''}%
\providecommand \bibnamefont  [1]{#1}%
\providecommand \bibfnamefont [1]{#1}%
\providecommand \citenamefont [1]{#1}%
\providecommand \href@noop [0]{\@secondoftwo}%
\providecommand \href [0]{\begingroup \@sanitize@url \@href}%
\providecommand \@href[1]{\@@startlink{#1}\@@href}%
\providecommand \@@href[1]{\endgroup#1\@@endlink}%
\providecommand \@sanitize@url [0]{\catcode `\\12\catcode `\$12\catcode
  `\&12\catcode `\#12\catcode `\^12\catcode `\_12\catcode `\%12\relax}%
\providecommand \@@startlink[1]{}%
\providecommand \@@endlink[0]{}%
\providecommand \url  [0]{\begingroup\@sanitize@url \@url }%
\providecommand \@url [1]{\endgroup\@href {#1}{\urlprefix }}%
\providecommand \urlprefix  [0]{URL }%
\providecommand \Eprint [0]{\href }%
\providecommand \doibase [0]{http://dx.doi.org/}%
\providecommand \selectlanguage [0]{\@gobble}%
\providecommand \bibinfo  [0]{\@secondoftwo}%
\providecommand \bibfield  [0]{\@secondoftwo}%
\providecommand \translation [1]{[#1]}%
\providecommand \BibitemOpen [0]{}%
\providecommand \bibitemStop [0]{}%
\providecommand \bibitemNoStop [0]{.\EOS\space}%
\providecommand \EOS [0]{\spacefactor3000\relax}%
\providecommand \BibitemShut  [1]{\csname bibitem#1\endcsname}%
\let\auto@bib@innerbib\@empty
%</preamble>
\bibitem [{\citenamefont {Moore}\ and\ \citenamefont {Read}(1991)}]{moore1991}%
  \BibitemOpen
  \bibfield  {author} {\bibinfo {author} {\bibfnamefont {G.}~\bibnamefont
  {Moore}}\ and\ \bibinfo {author} {\bibfnamefont {N.}~\bibnamefont {Read}},\
  }\href@noop {} {\bibfield  {journal} {\bibinfo  {journal} {Nucl. Phys. B}\
  }\textbf {\bibinfo {volume} {360}},\ \bibinfo {pages} {362} (\bibinfo {year}
  {1991})}\BibitemShut {NoStop}%
\bibitem [{\citenamefont {Read}\ and\ \citenamefont {Green}(2000)}]{read2000}%
  \BibitemOpen
  \bibfield  {author} {\bibinfo {author} {\bibfnamefont {N.}~\bibnamefont
  {Read}}\ and\ \bibinfo {author} {\bibfnamefont {D.}~\bibnamefont {Green}},\
  }\href {\doibase 10.1103/PhysRevB.61.10267} {\bibfield  {journal} {\bibinfo
  {journal} {Phys. Rev. B}\ }\textbf {\bibinfo {volume} {61}},\ \bibinfo
  {pages} {10267} (\bibinfo {year} {2000})}\BibitemShut {NoStop}%
\bibitem [{\citenamefont {Ivanov}(2001)}]{ivanov2001}%
  \BibitemOpen
  \bibfield  {author} {\bibinfo {author} {\bibfnamefont {D.~A.}\ \bibnamefont
  {Ivanov}},\ }\href {\doibase 10.1103/PhysRevLett.86.268} {\bibfield
  {journal} {\bibinfo  {journal} {Phys. Rev. Lett.}\ }\textbf {\bibinfo
  {volume} {86}},\ \bibinfo {pages} {268} (\bibinfo {year} {2001})}\BibitemShut
  {NoStop}%
\bibitem [{\citenamefont {Mackenzie}\ and\ \citenamefont
  {Maeno}(2003)}]{mackenzie2003}%
  \BibitemOpen
  \bibfield  {author} {\bibinfo {author} {\bibfnamefont {A.~P.}\ \bibnamefont
  {Mackenzie}}\ and\ \bibinfo {author} {\bibfnamefont {Y.}~\bibnamefont
  {Maeno}},\ }\href {\doibase 10.1103/RevModPhys.75.657} {\bibfield  {journal}
  {\bibinfo  {journal} {Rev. Mod. Phys.}\ }\textbf {\bibinfo {volume} {75}},\
  \bibinfo {pages} {657} (\bibinfo {year} {2003})}\BibitemShut {NoStop}%
\bibitem [{\citenamefont {Kitaev}(2006)}]{kitaev2006}%
  \BibitemOpen
  \bibfield  {author} {\bibinfo {author} {\bibfnamefont {A.}~\bibnamefont
  {Kitaev}},\ }\href {\doibase http://doi.org/10.1016/j.aop.2005.10.005}
  {\bibfield  {journal} {\bibinfo  {journal} {Ann. Phys.}\ }\textbf {\bibinfo
  {volume} {321}},\ \bibinfo {pages} {2} (\bibinfo {year} {2006})}\BibitemShut
  {NoStop}%
\bibitem [{\citenamefont {Fu}\ and\ \citenamefont {Kane}(2008)}]{fu2008}%
  \BibitemOpen
  \bibfield  {author} {\bibinfo {author} {\bibfnamefont {L.}~\bibnamefont
  {Fu}}\ and\ \bibinfo {author} {\bibfnamefont {C.~L.}\ \bibnamefont {Kane}},\
  }\href {\doibase 10.1103/PhysRevLett.100.096407} {\bibfield  {journal}
  {\bibinfo  {journal} {Phys. Rev. Lett.}\ }\textbf {\bibinfo {volume} {100}},\
  \bibinfo {pages} {096407} (\bibinfo {year} {2008})}\BibitemShut {NoStop}%
\bibitem [{\citenamefont {Sau}\ \emph {et~al.}(2010)\citenamefont {Sau},
  \citenamefont {Lutchyn}, \citenamefont {Tewari},\ and\ \citenamefont
  {Das~Sarma}}]{sau2010}%
  \BibitemOpen
  \bibfield  {author} {\bibinfo {author} {\bibfnamefont {J.~D.}\ \bibnamefont
  {Sau}}, \bibinfo {author} {\bibfnamefont {R.~M.}\ \bibnamefont {Lutchyn}},
  \bibinfo {author} {\bibfnamefont {S.}~\bibnamefont {Tewari}}, \ and\ \bibinfo
  {author} {\bibfnamefont {S.}~\bibnamefont {Das~Sarma}},\ }\href {\doibase
  10.1103/PhysRevLett.104.040502} {\bibfield  {journal} {\bibinfo  {journal}
  {Phys. Rev. Lett.}\ }\textbf {\bibinfo {volume} {104}},\ \bibinfo {pages}
  {040502} (\bibinfo {year} {2010})}\BibitemShut {NoStop}%
\bibitem [{\citenamefont {Alicea}(2010)}]{alicea2010}%
  \BibitemOpen
  \bibfield  {author} {\bibinfo {author} {\bibfnamefont {J.}~\bibnamefont
  {Alicea}},\ }\href {\doibase 10.1103/PhysRevB.81.125318} {\bibfield
  {journal} {\bibinfo  {journal} {Phys. Rev. B}\ }\textbf {\bibinfo {volume}
  {81}},\ \bibinfo {pages} {125318} (\bibinfo {year} {2010})}\BibitemShut
  {NoStop}%
\bibitem [{\citenamefont {Qi}\ \emph {et~al.}(2009)\citenamefont {Qi},
  \citenamefont {Hughes}, \citenamefont {Raghu},\ and\ \citenamefont
  {Zhang}}]{qi2009}%
  \BibitemOpen
  \bibfield  {author} {\bibinfo {author} {\bibfnamefont {X.-L.}\ \bibnamefont
  {Qi}}, \bibinfo {author} {\bibfnamefont {T.~L.}\ \bibnamefont {Hughes}},
  \bibinfo {author} {\bibfnamefont {S.}~\bibnamefont {Raghu}}, \ and\ \bibinfo
  {author} {\bibfnamefont {S.-C.}\ \bibnamefont {Zhang}},\ }\href {\doibase
  10.1103/PhysRevLett.102.187001} {\bibfield  {journal} {\bibinfo  {journal}
  {Phys. Rev. Lett.}\ }\textbf {\bibinfo {volume} {102}},\ \bibinfo {pages}
  {187001} (\bibinfo {year} {2009})}\BibitemShut {NoStop}%
\bibitem [{\citenamefont {Schnyder}\ \emph {et~al.}(2008)\citenamefont
  {Schnyder}, \citenamefont {Ryu}, \citenamefont {Furusaki},\ and\
  \citenamefont {Ludwig}}]{schnyder2008}%
  \BibitemOpen
  \bibfield  {author} {\bibinfo {author} {\bibfnamefont {A.~P.}\ \bibnamefont
  {Schnyder}}, \bibinfo {author} {\bibfnamefont {S.}~\bibnamefont {Ryu}},
  \bibinfo {author} {\bibfnamefont {A.}~\bibnamefont {Furusaki}}, \ and\
  \bibinfo {author} {\bibfnamefont {A.~W.~W.}\ \bibnamefont {Ludwig}},\ }\href
  {\doibase 10.1103/PhysRevB.78.195125} {\bibfield  {journal} {\bibinfo
  {journal} {Phys. Rev. B}\ }\textbf {\bibinfo {volume} {78}},\ \bibinfo
  {pages} {195125} (\bibinfo {year} {2008})}\BibitemShut {NoStop}%
\bibitem [{\citenamefont {Kitaev}(2009)}]{kitaev2009}%
  \BibitemOpen
  \bibfield  {author} {\bibinfo {author} {\bibfnamefont {A.}~\bibnamefont
  {Kitaev}},\ }\href@noop {} {\bibfield  {journal} {\bibinfo  {journal} {AIP
  Conf. Proc.}\ }\textbf {\bibinfo {volume} {1134}},\ \bibinfo {pages} {22}
  (\bibinfo {year} {2009})}\BibitemShut {NoStop}%
\bibitem [{\citenamefont {He}\ \emph {et~al.}(2017)\citenamefont {He},
  \citenamefont {Pan}, \citenamefont {Stern}, \citenamefont {Burks},
  \citenamefont {Che}, \citenamefont {Yin}, \citenamefont {Wang}, \citenamefont
  {Lian}, \citenamefont {Zhou}, \citenamefont {Choi}, \citenamefont {Murata},
  \citenamefont {Kou}, \citenamefont {Chen}, \citenamefont {Nie}, \citenamefont
  {Shao}, \citenamefont {Fan}, \citenamefont {Zhang}, \citenamefont {Liu},
  \citenamefont {Xia},\ and\ \citenamefont {Wang}}]{he2017}%
  \BibitemOpen
  \bibfield  {author} {\bibinfo {author} {\bibfnamefont {Q.~L.}\ \bibnamefont
  {He}}, \bibinfo {author} {\bibfnamefont {L.}~\bibnamefont {Pan}}, \bibinfo
  {author} {\bibfnamefont {A.~L.}\ \bibnamefont {Stern}}, \bibinfo {author}
  {\bibfnamefont {E.~C.}\ \bibnamefont {Burks}}, \bibinfo {author}
  {\bibfnamefont {X.}~\bibnamefont {Che}}, \bibinfo {author} {\bibfnamefont
  {G.}~\bibnamefont {Yin}}, \bibinfo {author} {\bibfnamefont {J.}~\bibnamefont
  {Wang}}, \bibinfo {author} {\bibfnamefont {B.}~\bibnamefont {Lian}}, \bibinfo
  {author} {\bibfnamefont {Q.}~\bibnamefont {Zhou}}, \bibinfo {author}
  {\bibfnamefont {E.~S.}\ \bibnamefont {Choi}}, \bibinfo {author}
  {\bibfnamefont {K.}~\bibnamefont {Murata}}, \bibinfo {author} {\bibfnamefont
  {X.}~\bibnamefont {Kou}}, \bibinfo {author} {\bibfnamefont {Z.}~\bibnamefont
  {Chen}}, \bibinfo {author} {\bibfnamefont {T.}~\bibnamefont {Nie}}, \bibinfo
  {author} {\bibfnamefont {Q.}~\bibnamefont {Shao}}, \bibinfo {author}
  {\bibfnamefont {Y.}~\bibnamefont {Fan}}, \bibinfo {author} {\bibfnamefont
  {S.-C.}\ \bibnamefont {Zhang}}, \bibinfo {author} {\bibfnamefont
  {K.}~\bibnamefont {Liu}}, \bibinfo {author} {\bibfnamefont {J.}~\bibnamefont
  {Xia}}, \ and\ \bibinfo {author} {\bibfnamefont {K.~L.}\ \bibnamefont
  {Wang}},\ }\href {\doibase 10.1126/science.aag2792} {\bibfield  {journal}
  {\bibinfo  {journal} {Science}\ }\textbf {\bibinfo {volume} {357}},\ \bibinfo
  {pages} {294} (\bibinfo {year} {2017})}\BibitemShut {NoStop}%
\bibitem [{\citenamefont {Williams}\ \emph {et~al.}(2007)\citenamefont
  {Williams}, \citenamefont {DiCarlo},\ and\ \citenamefont
  {Marcus}}]{williams2007}%
  \BibitemOpen
  \bibfield  {author} {\bibinfo {author} {\bibfnamefont {J.~R.}\ \bibnamefont
  {Williams}}, \bibinfo {author} {\bibfnamefont {L.}~\bibnamefont {DiCarlo}}, \
  and\ \bibinfo {author} {\bibfnamefont {C.~M.}\ \bibnamefont {Marcus}},\
  }\href {\doibase 10.1126/science.1144657} {\bibfield  {journal} {\bibinfo
  {journal} {Science}\ }\textbf {\bibinfo {volume} {317}},\ \bibinfo {pages}
  {638} (\bibinfo {year} {2007})}\BibitemShut {NoStop}%
\bibitem [{\citenamefont {\"Ozyilmaz}\ \emph {et~al.}(2007)\citenamefont
  {\"Ozyilmaz}, \citenamefont {Jarillo-Herrero}, \citenamefont {Efetov},
  \citenamefont {Abanin}, \citenamefont {Levitov},\ and\ \citenamefont
  {Kim}}]{kim2007}%
  \BibitemOpen
  \bibfield  {author} {\bibinfo {author} {\bibfnamefont {B.}~\bibnamefont
  {\"Ozyilmaz}}, \bibinfo {author} {\bibfnamefont {P.}~\bibnamefont
  {Jarillo-Herrero}}, \bibinfo {author} {\bibfnamefont {D.}~\bibnamefont
  {Efetov}}, \bibinfo {author} {\bibfnamefont {D.~A.}\ \bibnamefont {Abanin}},
  \bibinfo {author} {\bibfnamefont {L.~S.}\ \bibnamefont {Levitov}}, \ and\
  \bibinfo {author} {\bibfnamefont {P.}~\bibnamefont {Kim}},\ }\href {\doibase
  10.1103/PhysRevLett.99.166804} {\bibfield  {journal} {\bibinfo  {journal}
  {Phys. Rev. Lett.}\ }\textbf {\bibinfo {volume} {99}},\ \bibinfo {pages}
  {166804} (\bibinfo {year} {2007})}\BibitemShut {NoStop}%
\bibitem [{\citenamefont {Amet}\ \emph {et~al.}(2014)\citenamefont {Amet},
  \citenamefont {Williams}, \citenamefont {Watanabe}, \citenamefont
  {Taniguchi},\ and\ \citenamefont {Goldhaber-Gordon}}]{amet2014}%
  \BibitemOpen
  \bibfield  {author} {\bibinfo {author} {\bibfnamefont {F.}~\bibnamefont
  {Amet}}, \bibinfo {author} {\bibfnamefont {J.~R.}\ \bibnamefont {Williams}},
  \bibinfo {author} {\bibfnamefont {K.}~\bibnamefont {Watanabe}}, \bibinfo
  {author} {\bibfnamefont {T.}~\bibnamefont {Taniguchi}}, \ and\ \bibinfo
  {author} {\bibfnamefont {D.}~\bibnamefont {Goldhaber-Gordon}},\ }\href
  {\doibase 10.1103/PhysRevLett.112.196601} {\bibfield  {journal} {\bibinfo
  {journal} {Phys. Rev. Lett.}\ }\textbf {\bibinfo {volume} {112}},\ \bibinfo
  {pages} {196601} (\bibinfo {year} {2014})}\BibitemShut {NoStop}%
\bibitem [{\citenamefont {Qi}\ \emph {et~al.}(2010)\citenamefont {Qi},
  \citenamefont {Hughes},\ and\ \citenamefont {Zhang}}]{qi2010b}%
  \BibitemOpen
  \bibfield  {author} {\bibinfo {author} {\bibfnamefont {X.-L.}\ \bibnamefont
  {Qi}}, \bibinfo {author} {\bibfnamefont {T.~L.}\ \bibnamefont {Hughes}}, \
  and\ \bibinfo {author} {\bibfnamefont {S.-C.}\ \bibnamefont {Zhang}},\ }\href
  {\doibase 10.1103/PhysRevB.82.184516} {\bibfield  {journal} {\bibinfo
  {journal} {Phys. Rev. B}\ }\textbf {\bibinfo {volume} {82}},\ \bibinfo
  {pages} {184516} (\bibinfo {year} {2010})}\BibitemShut {NoStop}%
\bibitem [{\citenamefont {Chung}\ \emph {et~al.}(2011)\citenamefont {Chung},
  \citenamefont {Qi}, \citenamefont {Maciejko},\ and\ \citenamefont
  {Zhang}}]{chung2011}%
  \BibitemOpen
  \bibfield  {author} {\bibinfo {author} {\bibfnamefont {S.~B.}\ \bibnamefont
  {Chung}}, \bibinfo {author} {\bibfnamefont {X.-L.}\ \bibnamefont {Qi}},
  \bibinfo {author} {\bibfnamefont {J.}~\bibnamefont {Maciejko}}, \ and\
  \bibinfo {author} {\bibfnamefont {S.-C.}\ \bibnamefont {Zhang}},\ }\href
  {\doibase 10.1103/PhysRevB.83.100512} {\bibfield  {journal} {\bibinfo
  {journal} {Phys. Rev. B}\ }\textbf {\bibinfo {volume} {83}},\ \bibinfo
  {pages} {100512} (\bibinfo {year} {2011})}\BibitemShut {NoStop}%
\bibitem [{\citenamefont {Wang}\ \emph {et~al.}(2015)\citenamefont {Wang},
  \citenamefont {Zhou}, \citenamefont {Lian},\ and\ \citenamefont
  {Zhang}}]{wang2015c}%
  \BibitemOpen
  \bibfield  {author} {\bibinfo {author} {\bibfnamefont {J.}~\bibnamefont
  {Wang}}, \bibinfo {author} {\bibfnamefont {Q.}~\bibnamefont {Zhou}}, \bibinfo
  {author} {\bibfnamefont {B.}~\bibnamefont {Lian}}, \ and\ \bibinfo {author}
  {\bibfnamefont {S.-C.}\ \bibnamefont {Zhang}},\ }\href {\doibase
  10.1103/PhysRevB.92.064520} {\bibfield  {journal} {\bibinfo  {journal} {Phys.
  Rev. B}\ }\textbf {\bibinfo {volume} {92}},\ \bibinfo {pages} {064520}
  (\bibinfo {year} {2015})}\BibitemShut {NoStop}%
\bibitem [{\citenamefont {Lian}\ \emph {et~al.}()\citenamefont {Lian},
  \citenamefont {Sun}, \citenamefont {Vaezi}, \citenamefont {Qi},\ and\
  \citenamefont {Zhang}}]{lian2017}%
  \BibitemOpen
  \bibfield  {author} {\bibinfo {author} {\bibfnamefont {B.}~\bibnamefont
  {Lian}}, \bibinfo {author} {\bibfnamefont {X.-Q.}\ \bibnamefont {Sun}},
  \bibinfo {author} {\bibfnamefont {A.}~\bibnamefont {Vaezi}}, \bibinfo
  {author} {\bibfnamefont {X.-L.}\ \bibnamefont {Qi}}, \ and\ \bibinfo {author}
  {\bibfnamefont {S.-C.}\ \bibnamefont {Zhang}},\ }\href@noop {} {}\bibinfo
  {howpublished} {arXiv:1712.06156 (2017)}\BibitemShut {NoStop}%
\bibitem [{\citenamefont {Lian}\ \emph {et~al.}(2018)\citenamefont {Lian},
  \citenamefont {Wang}, \citenamefont {Sun}, \citenamefont {Vaezi},\ and\
  \citenamefont {Zhang}}]{lian2018}%
  \BibitemOpen
  \bibfield  {author} {\bibinfo {author} {\bibfnamefont {B.}~\bibnamefont
  {Lian}}, \bibinfo {author} {\bibfnamefont {J.}~\bibnamefont {Wang}}, \bibinfo
  {author} {\bibfnamefont {X.-Q.}\ \bibnamefont {Sun}}, \bibinfo {author}
  {\bibfnamefont {A.}~\bibnamefont {Vaezi}}, \ and\ \bibinfo {author}
  {\bibfnamefont {S.-C.}\ \bibnamefont {Zhang}},\ }\href {\doibase
  10.1103/PhysRevB.97.125408} {\bibfield  {journal} {\bibinfo  {journal} {Phys.
  Rev. B}\ }\textbf {\bibinfo {volume} {97}},\ \bibinfo {pages} {125408}
  (\bibinfo {year} {2018})}\BibitemShut {NoStop}%
\bibitem [{\citenamefont {Anantram}\ and\ \citenamefont
  {Datta}(1996)}]{anantram1996}%
  \BibitemOpen
  \bibfield  {author} {\bibinfo {author} {\bibfnamefont {M.~P.}\ \bibnamefont
  {Anantram}}\ and\ \bibinfo {author} {\bibfnamefont {S.}~\bibnamefont
  {Datta}},\ }\href {\doibase 10.1103/PhysRevB.53.16390} {\bibfield  {journal}
  {\bibinfo  {journal} {Phys. Rev. B}\ }\textbf {\bibinfo {volume} {53}},\
  \bibinfo {pages} {16390} (\bibinfo {year} {1996})}\BibitemShut {NoStop}%
\bibitem [{\citenamefont {Entin-Wohlman}\ \emph {et~al.}(2008)\citenamefont
  {Entin-Wohlman}, \citenamefont {Imry},\ and\ \citenamefont
  {Aharony}}]{entin2008}%
  \BibitemOpen
  \bibfield  {author} {\bibinfo {author} {\bibfnamefont {O.}~\bibnamefont
  {Entin-Wohlman}}, \bibinfo {author} {\bibfnamefont {Y.}~\bibnamefont {Imry}},
  \ and\ \bibinfo {author} {\bibfnamefont {A.}~\bibnamefont {Aharony}},\ }\href
  {\doibase 10.1103/PhysRevB.78.224510} {\bibfield  {journal} {\bibinfo
  {journal} {Phys. Rev. B}\ }\textbf {\bibinfo {volume} {78}},\ \bibinfo
  {pages} {224510} (\bibinfo {year} {2008})}\BibitemShut {NoStop}%
\bibitem [{sup()}]{supple}%
  \BibitemOpen
  \href@noop {} {}\bibinfo {note} {See Supplemental Material for technical
  details.}\BibitemShut {Stop}%
\bibitem [{\citenamefont {Levin}\ \emph {et~al.}(2007)\citenamefont {Levin},
  \citenamefont {Halperin},\ and\ \citenamefont {Rosenow}}]{levin2007}%
  \BibitemOpen
  \bibfield  {author} {\bibinfo {author} {\bibfnamefont {M.}~\bibnamefont
  {Levin}}, \bibinfo {author} {\bibfnamefont {B.~I.}\ \bibnamefont {Halperin}},
  \ and\ \bibinfo {author} {\bibfnamefont {B.}~\bibnamefont {Rosenow}},\ }\href
  {\doibase 10.1103/PhysRevLett.99.236806} {\bibfield  {journal} {\bibinfo
  {journal} {Phys. Rev. Lett.}\ }\textbf {\bibinfo {volume} {99}},\ \bibinfo
  {pages} {236806} (\bibinfo {year} {2007})}\BibitemShut {NoStop}%
\bibitem [{\citenamefont {Lee}\ \emph {et~al.}(2007)\citenamefont {Lee},
  \citenamefont {Ryu}, \citenamefont {Nayak},\ and\ \citenamefont
  {Fisher}}]{lee2007}%
  \BibitemOpen
  \bibfield  {author} {\bibinfo {author} {\bibfnamefont {S.-S.}\ \bibnamefont
  {Lee}}, \bibinfo {author} {\bibfnamefont {S.}~\bibnamefont {Ryu}}, \bibinfo
  {author} {\bibfnamefont {C.}~\bibnamefont {Nayak}}, \ and\ \bibinfo {author}
  {\bibfnamefont {M.~P.~A.}\ \bibnamefont {Fisher}},\ }\href {\doibase
  10.1103/PhysRevLett.99.236807} {\bibfield  {journal} {\bibinfo  {journal}
  {Phys. Rev. Lett.}\ }\textbf {\bibinfo {volume} {99}},\ \bibinfo {pages}
  {236807} (\bibinfo {year} {2007})}\BibitemShut {NoStop}%
\bibitem [{\citenamefont {Lian}\ and\ \citenamefont {Wang}(2018)}]{lian2018a}%
  \BibitemOpen
  \bibfield  {author} {\bibinfo {author} {\bibfnamefont {B.}~\bibnamefont
  {Lian}}\ and\ \bibinfo {author} {\bibfnamefont {J.}~\bibnamefont {Wang}},\
  }\href {\doibase 10.1103/PhysRevB.97.165124} {\bibfield  {journal} {\bibinfo
  {journal} {Phys. Rev. B}\ }\textbf {\bibinfo {volume} {97}},\ \bibinfo
  {pages} {165124} (\bibinfo {year} {2018})}\BibitemShut {NoStop}%
\bibitem [{\citenamefont {Lian}\ \emph {et~al.}(2016)\citenamefont {Lian},
  \citenamefont {Wang},\ and\ \citenamefont {Zhang}}]{lian2016a}%
  \BibitemOpen
  \bibfield  {author} {\bibinfo {author} {\bibfnamefont {B.}~\bibnamefont
  {Lian}}, \bibinfo {author} {\bibfnamefont {J.}~\bibnamefont {Wang}}, \ and\
  \bibinfo {author} {\bibfnamefont {S.-C.}\ \bibnamefont {Zhang}},\ }\href
  {\doibase 10.1103/PhysRevB.93.161401} {\bibfield  {journal} {\bibinfo
  {journal} {Phys. Rev. B}\ }\textbf {\bibinfo {volume} {93}},\ \bibinfo
  {pages} {161401} (\bibinfo {year} {2016})}\BibitemShut {NoStop}%
\bibitem [{\citenamefont {Banerjee}\ \emph {et~al.}(2017)\citenamefont
  {Banerjee}, \citenamefont {Heiblum}, \citenamefont {Rosenblatt},
  \citenamefont {Oreg}, \citenamefont {Feldman}, \citenamefont {Stern},\ and\
  \citenamefont {Umansky}}]{banerjee2017}%
  \BibitemOpen
  \bibfield  {author} {\bibinfo {author} {\bibfnamefont {M.}~\bibnamefont
  {Banerjee}}, \bibinfo {author} {\bibfnamefont {M.}~\bibnamefont {Heiblum}},
  \bibinfo {author} {\bibfnamefont {A.}~\bibnamefont {Rosenblatt}}, \bibinfo
  {author} {\bibfnamefont {Y.}~\bibnamefont {Oreg}}, \bibinfo {author}
  {\bibfnamefont {D.~E.}\ \bibnamefont {Feldman}}, \bibinfo {author}
  {\bibfnamefont {A.}~\bibnamefont {Stern}}, \ and\ \bibinfo {author}
  {\bibfnamefont {V.}~\bibnamefont {Umansky}},\ }\href@noop {} {\bibfield
  {journal} {\bibinfo  {journal} {Nature}\ }\textbf {\bibinfo {volume} {545}},\
  \bibinfo {pages} {75} (\bibinfo {year} {2017})}\BibitemShut {NoStop}%
\bibitem [{\citenamefont {Zhang}\ \emph {et~al.}(2018)\citenamefont {Zhang},
  \citenamefont {Yaji}, \citenamefont {Hashimoto}, \citenamefont {Ota},
  \citenamefont {Kondo}, \citenamefont {Okazaki}, \citenamefont {Wang},
  \citenamefont {Wen}, \citenamefont {Gu}, \citenamefont {Ding},\ and\
  \citenamefont {Shin}}]{zhang2018}%
  \BibitemOpen
  \bibfield  {author} {\bibinfo {author} {\bibfnamefont {P.}~\bibnamefont
  {Zhang}}, \bibinfo {author} {\bibfnamefont {K.}~\bibnamefont {Yaji}},
  \bibinfo {author} {\bibfnamefont {T.}~\bibnamefont {Hashimoto}}, \bibinfo
  {author} {\bibfnamefont {Y.}~\bibnamefont {Ota}}, \bibinfo {author}
  {\bibfnamefont {T.}~\bibnamefont {Kondo}}, \bibinfo {author} {\bibfnamefont
  {K.}~\bibnamefont {Okazaki}}, \bibinfo {author} {\bibfnamefont
  {Z.}~\bibnamefont {Wang}}, \bibinfo {author} {\bibfnamefont {J.}~\bibnamefont
  {Wen}}, \bibinfo {author} {\bibfnamefont {G.~D.}\ \bibnamefont {Gu}},
  \bibinfo {author} {\bibfnamefont {H.}~\bibnamefont {Ding}}, \ and\ \bibinfo
  {author} {\bibfnamefont {S.}~\bibnamefont {Shin}},\ }\href {\doibase
  10.1126/science.aan4596} {\bibfield  {journal} {\bibinfo  {journal}
  {Science}\ } (\bibinfo {year} {2018}),\ 10.1126/science.aan4596}\BibitemShut
  {NoStop}%
\bibitem [{\citenamefont {Jiang}\ \emph {et~al.}(2018)\citenamefont {Jiang},
  \citenamefont {Feng}, \citenamefont {Wu}, \citenamefont {Li}, \citenamefont
  {Bai}, \citenamefont {Li}, \citenamefont {Zhang}, \citenamefont {Gu},
  \citenamefont {Feng}, \citenamefont {Zhang}, \citenamefont {Song},
  \citenamefont {Wang}, \citenamefont {Li}, \citenamefont {Ma}, \citenamefont
  {Xue}, \citenamefont {Wang},\ and\ \citenamefont {He}}]{jiang2018}%
  \BibitemOpen
  \bibfield  {author} {\bibinfo {author} {\bibfnamefont {G.}~\bibnamefont
  {Jiang}}, \bibinfo {author} {\bibfnamefont {Y.}~\bibnamefont {Feng}},
  \bibinfo {author} {\bibfnamefont {W.}~\bibnamefont {Wu}}, \bibinfo {author}
  {\bibfnamefont {S.}~\bibnamefont {Li}}, \bibinfo {author} {\bibfnamefont
  {Y.}~\bibnamefont {Bai}}, \bibinfo {author} {\bibfnamefont {Y.}~\bibnamefont
  {Li}}, \bibinfo {author} {\bibfnamefont {Q.}~\bibnamefont {Zhang}}, \bibinfo
  {author} {\bibfnamefont {L.}~\bibnamefont {Gu}}, \bibinfo {author}
  {\bibfnamefont {X.}~\bibnamefont {Feng}}, \bibinfo {author} {\bibfnamefont
  {D.}~\bibnamefont {Zhang}}, \bibinfo {author} {\bibfnamefont
  {C.}~\bibnamefont {Song}}, \bibinfo {author} {\bibfnamefont {L.}~\bibnamefont
  {Wang}}, \bibinfo {author} {\bibfnamefont {W.}~\bibnamefont {Li}}, \bibinfo
  {author} {\bibfnamefont {X.-C.}\ \bibnamefont {Ma}}, \bibinfo {author}
  {\bibfnamefont {Q.-K.}\ \bibnamefont {Xue}}, \bibinfo {author} {\bibfnamefont
  {Y.}~\bibnamefont {Wang}}, \ and\ \bibinfo {author} {\bibfnamefont
  {K.}~\bibnamefont {He}},\ }\href
  {http://stacks.iop.org/0256-307X/35/i=7/a=076802} {\bibfield  {journal}
  {\bibinfo  {journal} {Chinese Physics Letters}\ }\textbf {\bibinfo {volume}
  {35}},\ \bibinfo {pages} {076802} (\bibinfo {year} {2018})}\BibitemShut
  {NoStop}%
\bibitem [{\citenamefont {{Wang}}\ and\ \citenamefont
  {{Lian}}(2018)}]{wang2018}%
  \BibitemOpen
  \bibfield  {author} {\bibinfo {author} {\bibfnamefont {J.}~\bibnamefont
  {{Wang}}}\ and\ \bibinfo {author} {\bibfnamefont {B.}~\bibnamefont
  {{Lian}}},\ }\href@noop {} {\bibfield  {journal} {\bibinfo  {journal} {ArXiv
  e-prints}\ } (\bibinfo {year} {2018})},\ \Eprint
  {http://arxiv.org/abs/1805.10763} {arXiv:1805.10763 [cond-mat.mes-hall]}
  \BibitemShut {NoStop}%
\end{thebibliography}%

\end{document}